\documentclass[twocolumn,showpacs,preprintnumbers,amsmath,amssymb,aps]{revtex4}
\usepackage{graphicx}
\begin{document}

\title{Positive and Negative Drag, Dynamic Phases, and Commensurability in Coupled One-Dimensional Channels of Particles with Yukawa Interactions}   
\author{
C. Reichhardt$^1$, 
C. Bairnsfather$^{1,2}$, 
and    
C. J. Olson Reichhardt$^1$} 
\affiliation{
$^1$Theoretical Division,
Los Alamos National Laboratory, Los Alamos, New Mexico 87545, USA\\
$^2$Department of Physics, Purdue University, West Lafayette, Indiana 47907, USA }

\date{\today}
\begin{abstract}
We introduce a simple model consisting of two or three coupled 
one-dimensional channels of particles
with Yukawa interactions. 
For the two channel system, when an external drive is applied only to the 
top or primary channel, we find a transition from locked flow
where particles in both channels move together 
to decoupled flow  where the particles in the secondary or undriven channel
move at a slower velocity than the particles in the primary or driven channel.
Pronounced commensurability effects in the decoupling transition occur when
the ratio of the number of particles in the top and bottom channels
is varied, and the coupling of the two channels is enhanced when this ratio is
an integer or a rational fraction.
Near the commensurate fillings, we find additional features in the
velocity-force curves caused by the slipping of individual vacancies 
or incommensurations in the secondary channels.
For three coupled channels, when only the top channel is driven
we find a remarkably rich variety of distinct dynamic phases,
including multiple decoupling and recoupling transitions.
These transitions produce
pronounced signatures in the velocity response of each channel.
We also find regimes where a negative drag effect can be 
induced in one of the non-driven channels.  The particles in this
channel move in the opposite direction from the particles in the driven
channel
due to the mixing of the two different periodic frequencies produced
by the discrete motion of the particles in the two other channels. 
In the two channel system, we also demonstrate a ratchet effect for the
particles in the secondary channel when an asymmetric drive is applied to
the primary channel.  This ratchet effect is
similar to that observed in superconducting vortex systems 
when there is a coupling between two different species of vortices.
\end{abstract}
\maketitle

\vskip2pc

\section{Introduction} 
There are many systems composed of repulsively interacting particles
with one dimensional (1D) or quasi-1D motion, including colloids 
in narrow channels 
\cite{Koppi,Mc,Doyl,Ferreira,Yang,Misko,Wei,Lutz,Bleil,Roichman},
Wigner crystal states in wires 
\cite{Sc,Glaz,R,Mueller,Bockrath,Hew,Peeters1,I,Meyer}  
and constrictions \cite{FPeeters}, dusty plasmas in grooves \cite{Goree}, 
macroscopic charged ball bearings in channels \cite{Coupier},
and vortices in type-II superconductors 
confined within narrow strips or channels \cite{Kes,K,A,Plourde,Reichhardt}. 
In many of these systems,
interesting structural transitions from 1D 
lines of particles to zig-zag or buckled states can occur 
\cite{Peeters1,Ferreira,Meyer}.  There can also be
higher order transitions 
from 2 rows to 3 rows of particles 
or transitions to disordered states \cite{Kes,K,Yang,Reichhardt}. 
Under an external drive, this type of system also exhibits a variety of 
dynamical behavior
such as ordered or disordered motion 
through constrictions \cite{Koppi,FPeeters,Reichhardt} 
or dynamic commensurability effects \cite{Kes,K}. 

\begin{figure}
\includegraphics[width=3.5in]{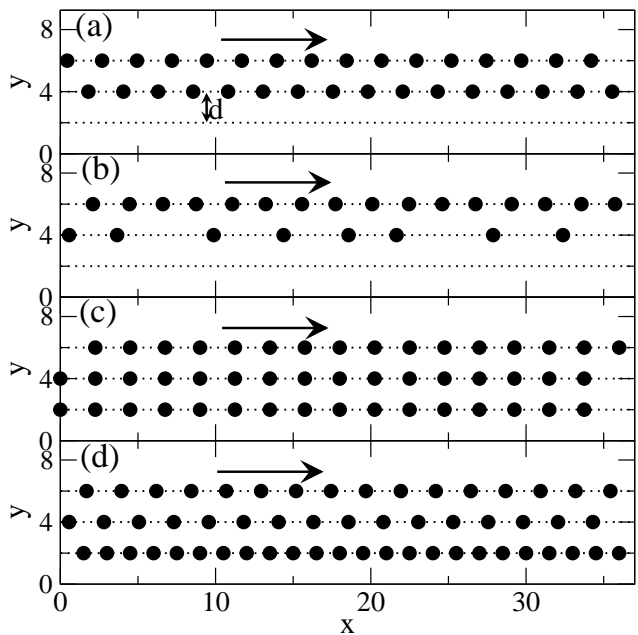}
\caption{
Image of the sample geometry.  The locations of the three channels are
indicated by dotted lines.  Black dots are particles within the channels.
The arrow denotes the driving force which is applied only to particles
within the top channel, termed the primary channel $p$.  The bottom 
undriven channels are the secondary channels $s_1$ and $s_2$.
The ratio of the number of particles in each channel is
$R_{s1,p}=N_{s1}/N_p$ and $R_{s2,p}=N_{s2}/N_p$, where $N_{s1}$ and $N_{s2}$ are
the number of particles in the secondary channels and $N_p=16$ is the number 
of particles in the primary channel.
The spacing between channels $d$ is marked in panel (a).
(a) Two channels with $R_{s1,p} = 1.0$ 
(b) Two channels with $R_{s1,p} = 0.5$. 
(c) Three channels with $R_{s1,p}=1.0$ and $R_{s2,p}=1.0$.
(d) Three channels with $R_{s1,p}=1.0$ and $R_{s2,p}= 1.5$.       
}
\label{fig:1}
\end{figure}

Here we propose a simple system consisting of particles in two 
or three coupled 1D channels. 
The particles in each channel interact with the other particles in the
same channel as well as with particles in adjacent channels 
via a Yukawa potential. 
An external drive applied to only one channel produces
drag effects on the particles in the undriven channels, causing
them to move.  Our system is illustrated in Fig.~\ref{fig:1}. 
The two channel system is similar to the transformer geometry 
studied for vortices in
two superconducting layers 
where an external drive is applied to one (primary) layer and the response
of the nondriven (secondary) layer is measured
\cite{Ig,Jr,Ekin}. 
If the vortices in the two layers
are fully coupled, the response of the secondary layer is
exactly the same as the response of the primary layer.
If the vortices are only partially coupled,
the response in the secondary layer is smaller than that of the primary
layer. 
The transformer geometry has also been studied for vortex systems
with multiple layers, such as vortices in the strongly 
layered high-temperature superconductors \cite{Pe}. 
Drag effects have also been predicted for two coupled 1D wires containing
classical 1D Wigner crystals when only one of the wires is driven \cite{R}. 
In this case the interaction between particles in neighboring wires 
is repulsive, unlike the attractive interaction between vortices in 
neighboring layers.
For the coupled 1D Wigner crystals there 
is also a transition from a completely locked state, where the response
in both wires is identical, to a partially locked state, where the response
of the secondary wire is reduced.
Drag effects in coupled wire experiments have been interpreted 
as arising from the formation of Wigner crystal states \cite{Stopa}. 

In this work we consider the effect of changing the ratio of the number of
particles in each channel on the locking or coupling between the channels,
with a particular focus on ratios that are integers or rational fractions.
Commensuration effects \cite{Bak} occur when the spacing between particles
in one channel is a simple rational fraction of the spacing between particles
in another channel, while incommensurations such as vacancies or interstitials
appear when the two spacings are incommensurable.
Commensuration effects
have been studied extensively for systems in which
a varied number of particles interacts with a rigid periodic substrate, 
such as atoms and molecules on surfaces \cite{Cm},
vortices in superconductors with periodic pinning arrays 
\cite{Baert,Nori,Met}, 
and colloids interacting with 1D \cite{Bechinger} or 
two-dimensional (2D) optical trap arrays \cite{Olson}, all of which
can be viewed as physical realizations of the Frenkel-Kontrova model.
These studies find that the coupling to the substrate or the effective pinning
of the particles by the substrate is strongly enhanced when the ratio of
the number of particles to the number of substrate minima is an integer or
a rational fraction, as indicated by the appearance of peaks in the critical
depinning force or enhanced ordering of the particles at the commensurate
fillings.
In our system, for the two channel geometry illustrated in Fig.~\ref{fig:1} 
the particles in the secondary channel can be regarded as a distortable or
moveable periodic pinning substrate for the particles 
in the primary channel, suggesting that enhanced drag or coupling 
could occur when the ratio of the number of particles in each channel is
an integer or a rational fraction.
The deformability of the substrate makes our proposed model distinct from
Frenkel-Kontrova systems.
Additionally, driven 1D and 2D commensurate-incommensurate systems  
often exhibit numerous dynamic behaviors within the
incommensurate regimes, such as when localized
vacancies or interstitials form soliton-like excitations 
which move more easily than the particles over the substrate
\cite{Periodic,Hu}. 
This suggests that similar phases may be possible in the 
coupled channel drag system 
we propose here, and we show that such phases do appear.
We also show that when we make the system more complex by adding a third
channel, a remarkable variety of commensuration effects and dynamic regimes 
occur such as multiple decoupling, recoupling, and slip transitions, all of
which produce 
pronounced changes in the velocity response. 
It is even possible to realize negative drag effects 
where the particles in one of the
channels move in the direction opposite to that of the applied drive.    

The coupled channels system we propose could be realized in 
colloidal systems.  The number of colloids in the different channels
can be controlled readily by optical manipulation and the colloids in
one channel could be driven with an external field, optically, or using
microfluidics.
Another possible realization of this system is in 
nanowires where 1D Wigner crystallization of the electrons has occurred; in
this case, by altering the electron density,
the particle lattice spacing in one wire could be 
varied with respect to that in an adjacent wire. 
Realizing such a system could have important implications 
for the study of 1D Wigner crystals since the
appearance of commensuration effects 
would be strong evidence that Wigner crystal states are forming. 
In superconducting systems, the density of magnetic vortices is 
fixed by the externally applied magnetic field, so it would be difficult to
create 1D channels that contain different linear densities of vortices;
however, in certain layered systems an additional transverse magnetic 
field can be applied to create a second Josephson vortex lattice 
which can interact with the pancake vortices in the planes 
\cite{Sav,Cole,Caplin}.  It has already been shown 
that using this technique it is possible
to drive only one of the vortex species and induce a drag on the other 
vortex species \cite{Cole,Caplin}. 
It should be possible to study fractional commensurate states 
in such a vortex system by examining how
the drag effect changes when the ratio of the number of one type of
vortices to the other is varied.
A realization of three or more channels with varied numbers of particles
in each channel should again be possible using colloidal systems or
metallic wires.  Further, a superconducting or nanowire system could be
used in which each layer or channel has the same number of particles but
differing amounts of quenched disorder.
We note that there are previous studies of colloidal particles 
in 2D bilayers \cite{HL} where the particles in the 
layers are driven in opposite directions;
however, these studies focused on an oscillatory order-disorder transition,
not on the effects of commensuration on decoupling or the dynamic phases 
that we consider here for the case of 1D coupled channels.

The paper is organized as follows: In Section II, we describe our simulation
method and sample geometry.  We consider two channels of particles in
Section III and illustrate
a drive-induced decoupling transition for commensurate channels
in Section III A. In Section III B we describe the two step
decoupling transition that occurs for incommensurate channels which
contain vacancy or interstitial sites that can act like a second species
of particle.  The effects of finite temperature and finite size appear
in Section III C.
Section III D shows that the nonlinear response of the system 
can be exploited to create a ratchet effect, where ac motion in the
driven channel induces dc transport in the drag channel.  
In Section IV we turn to samples with three channels.
We show in Section IV A that when the driven channel
is commensurate with the neighboring drag channel, four different types of 
coupled and decoupled flow can occur as the occupancy of the 
second drag channel is varied, including regimes of intermittent coupling.
In Section IV B, the driven channel is incommensurate with the neighboring drag
channel and we find a
complex series of coupling-decoupling transitions that
produce a significant amount of structure in the velocity-force curves.
In Section IV C, we consider in detail the negative drag that can occur
at incommensurate fillings 
when the particles in one of the drag channels move in the direction 
opposite to the particles in the driven channel.
It is also possible for the outer channels to remain coupled while the
central channel is decoupled, as described in Section IV D.
In Section IV E we summarize all five of the dynamical phases and the
negative drag
by showing that they can be achieved in a single system.
The paper concludes in Section V with a discussion and summary.

\section{Simulation}

We model interacting Yukawa particles confined to 
move along 1D channels as illustrated in 
Fig.~\ref{fig:1}.  
Each particle interacts with other particles in the same channel and
with particles in adjacent channels.
The separation between channels is $d=2$ and, unless otherwise noted, there are
$N_{p}=16$ particles in the driven or primary channel 
with a lattice spacing $a$, where $L$ is the length of the channel.
The particles in the primary channel $p$ are coupled to an applied 
external driving force 
$F_{D}$.  For a two channel system,
the drag or secondary channel $s_1$ contains $N_{s1}$ particles, 
and the commensurability ratio is $R_{s1,p} = N_{s1}/N_{p}$. 
In a three channel system such as that shown in Fig.~\ref{fig:1}(c,d),
the additional secondary channel $s_2$ is adjacent to $s_1$ but not to
the primary channel $p$, and it contains $N_{s2}$ particles, giving
a commensurability ratio of $R_{s2,p}=N_{s2}/N_p$.     

The particle motions evolve under 
overdamped dynamics where the colloids
obey the following equation of motion: 
\begin{equation} 
\eta \frac{d{\bf R}_i}{dt} = {\bf F}^{pp}_{i} + {\bf F}^{D}_{p}
\end{equation}
Here $\eta$ is the damping constant,
${\bf R}_{i}$ is the location of particle $i$, 
and the repulsive particle-particle interaction force is 
${\bf F}^{pp}_{i} = \sum^{N_{v}}_{j\ne i}-\nabla V(R_{ij})$. 
The potential has a Yukawa or screened Coulomb form 
of 
\begin{equation}
V(R_{ij}) = \frac{E_{0}}{R_{ij}}e^{-\kappa R_{ij}} 
\end{equation}
with 
$E_{0} = Z^{*2}/4\pi\epsilon\epsilon_{0}a_{0}$, where $\epsilon$ is the solvent  
dielectric constant, $Z^{*}$ is the effective charge, and $1/\kappa$ is the 
screening length.  
For colloidal systems, the length scale $a_{0}$ is on 
the order of a micron.  
We measure forces in units of $F_{0} = E_{0}/a_{0}$ and time in 
units of $\tau = \eta/E_{0}$.
In a typical case, the distance between particles in channel $p$ is
$a=2.25$ while the distance between adjacent channels is $d= 1.125a$ 
and the screening length is $1/\kappa=2d$, which is long enough
to ensure strong coupling between the particles in all three channels.
The driving force $F^{D}_p=F_D{\bf {\hat x}}$ 
is applied only to all the particles
in the primary channel. 
We increase $F_D$ from zero in small increments
of $\delta F_D$, holding the drive at constant
values for a fixed time interval during which we measure the velocity of
the particles in each channel.
We have carefully checked that our waiting times are long enough to eliminate
transient effects. 
We use $\delta F_{D} = 0.001$ and a wait time of $10^5$ simulation time steps.  
We impose periodic boundary conditions in the $x$-direction 
along the length of the channels.
The velocity of the particles in the primary channel is 
given by $V_{p}$ and the particle velocity in the
secondary channels is given by $V_{s1}$ and $V_{s2}$.
We normalize all 
the velocities by the number of particles in each channel,
$V_p=N_p^{-1}\sum_i^{N_p}{\bf v}_i$,
$V_{s1}=N_{s1}^{-1}\sum_i^{N_{s1}}{\bf v}_i$, and 
$V_{s2}=N_{s2}^{-1}\sum_i^{N_{s2}}{\bf v}_i$.

\begin{figure}
\includegraphics[width=3.5in]{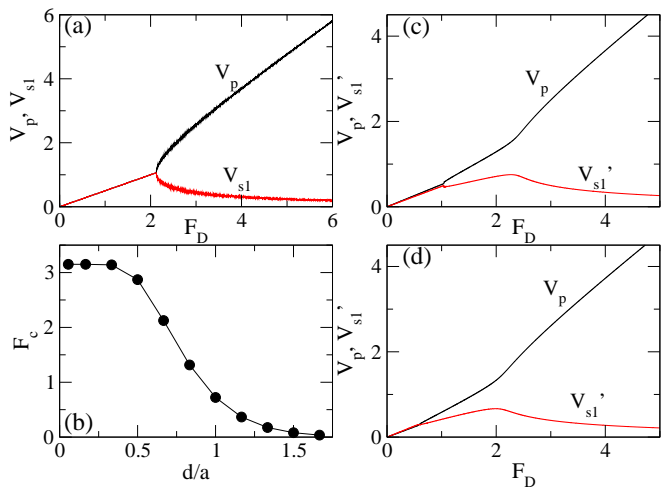}
\caption{
(a) The average velocity in the primary channel $V_{p}$ 
and the secondary channel $V_{s1}$ vs $F_{D}$ for a two channel system with
$R_{s1,p} = 1.0$ and $d/a = 0.67$. 
The channels are locked for low $F_D$ in the regime where $V_p$ and
$V_{s1}$ increase linearly.
At $F_{D} = 2.125$,
there is a transition to a partially decoupled state where 
the particles in the secondary channel begin to slip, producing
a decreasing $V_{s1}$. 
(b) The decoupling force $F_c$ force vs $d/a$ for the system in (a)
with $R_{s1,p}=1$, fixed $d = 2.0$, 
and $a$ altered by changing the particle
density. The decoupling transition drops to lower values of $F_D$
as the particle density increases.
(c) $V_{p}$ (upper curve) and $V_{s1}^{\prime}=N_p^{-1}\sum_i^{N_{s1}}{\bf v}_i$ 
(lower curve) vs $F_D$
for the system in (a) but with $R_{s1,p} = 0.92$. 
Unlike the commensurate case in (a), the initial 
unlocking phase above $F_D=1.04$ is associated with
the slipping of vacancies in $s_1$, while at $F_D=2.25$
there is a second unlocking transition above which
all the particles in $s_1$ slip with respect to the particles in $p$.
(d) $V_p$ (upper curve) and $V_{s1}^{\prime}$ (lower curve) vs $F_D$ for
$R_{s1,p} = 1.08$, where the slipping 
at low drives is due to the presence of incommensurate particles.
}
\label{fig:2}
\label{fig:3}
\label{fig:4}
\end{figure}

\section{Two Channel Systems}

\subsection{Coupling-Decoupling Transitions for Commensurate Channels}
We first focus on the two channel system at 
commensurate filling with a particle ratio of 
$R_{s1,p} = 1.0$ and with $d/a = 0.67$.  In Fig.~\ref{fig:2}(a)
we plot $V_{p}$ and $V_{s1}$ together versus $F_{D}$. 
Both $V_{p}$ and $V_{s1}$ increase linearly with $F_{D}$ for low $F_D$ and
have identical values, 
indicating that the motion in the two channels is locked.
At $F_{D}=2.125$, we find
a transition to a partially decoupled state 
where $V_{s1}$ monotonically decreases with increasing $F_{D}$ 
while $V_{p}$ continues to increase with $F_{D}$ at a rate faster than
the linear increase that occurred below the transition.
The particles in $s_1$ are not completely decoupled 
from the primary channel since they still exhibit a nonzero velocity; since
the particles in $s_1$ do not experience a driving force, they can move only
due to interactions with the particles in $p$.
Just above the decoupling transition, 
$V_{p}$ increases with a square root form. 
The general shape 
of the velocity force curves in Fig.~\ref{fig:2}(a) 
is the same as that of the current-voltage curves obtained for
superconducting transformer geometries \cite{Ig,Jr,Ekin,Pe}, 
where the vortex velocities are proportional
to the voltage and the applied current is 
proportional to the external force on the
vortices in the primary channel. 
The current-voltage curves in the superconducting transformer system 
indicate that there is 
a drive-induced decoupling transition of the vortices in adjacent layers.
In the vortex system, the vortex-vortex interaction 
between layers is attractive,
so it is more intuitive why a finite vortex mobility persists 
in the secondary channel above decoupling. 
The results in Fig.~\ref{fig:2}(a) indicate that
even when the particle-particle interactions are purely repulsive,
a finite velocity in the secondary channel can be maintained 
at drives above the decoupling transition. 

We measure the decoupling or unlocking force $F_c$ as a function of
$d/a$ in a two channel system with $R_{s1,p} =1.0$. 
The result is plotted in Fig.~\ref{fig:3}(b), where we fix $d=2.0$ and vary $a$ by
changing $N_p$ and $N_{s1}$.
Here $F_{c}$ decreases with increasing particle
density. A similar effect occurs in layered vortex systems, 
where for higher fields
or higher vortex densities the coupling between the 
layers is gradually reduced \cite{Pe}. For the
Yukawa system this effect can be attributed to
the reduced size of the periodic potential that the particles in $s_1$ 
experience from the particles in $p$. 
Once the primary channel is in motion, 
the particles in $s_1$ shift to positions that are slightly behind the
driven particles. 
As the drive increases, the size of this shift increases 
until the particles in $p$ slip more than $0.5a$ ahead of the
particles in $s_1$, producing
the partial decoupling. 
When the particle density increases, the amount of shift required to
pass the position $0.5a$ decreases since $a$ decreases with increasing
particle density in both channels.

\subsection{Dynamics and Commensurability}
In Fig.~\ref{fig:4}(c) we plot $V_{p}$ and $V_{s1}^{\prime}$ for 
a two channel system with $R_{s1,p} = 0.92$, where the number of particles
in $s_1$ is smaller than the number of particles in $p$.
Here $V_{s1}^{\prime}=N_p^{-1}\sum_i^{N_{s1}}{\bf v}_i$ is the velocity
of the particles in $s_1$ normalized by $N_p$, the number of particles in $p$.
For low $F_D$ the channels are locked and all of the particles
in the system move at the same velocity.
The slope of $V_{s1}^{\prime}$ versus $F_D$ is slightly smaller than the slope
of $V_p$ versus $F_D$ in Fig.~\ref{fig:4}(c) due to the fact that $N_{s1}<N_p$.
At $F_{D} = 1.04$ we observe a transition to a 
partially coupled state; however, this transition occurs
at a drive well below
the decoupling transition $F_c=2.125$ shown for the commensurate system
in Fig.~\ref{fig:2}(a).
Additionally, just above $F_D=1.04$, Fig.~\ref{fig:4}(c) indicates that
the velocity-force curve does not have 
the characteristic square root shape found close to $F_c$ 
in Fig.~\ref{fig:2}(a). 
For $F_D>1.04$,
$V_{s1}^{\prime}$ continues to increase with increasing
$F_{D}$ but with a smaller slope than in the locked regime. 
A second decoupling transition appears at $F_{D} = 2.25$. 
For $F_D>2.25$, $V_{s1}^{\prime}$ decreases with increasing $F_{D}$ and there is
also a corresponding increase in the slope of $V_p$.
The second decoupling transition occurs
at a drive close to the value $F_{c}=2.125$ 
where decoupling of the commensurate system
occurs, as shown in Fig.~\ref{fig:2}(a).  This indicates that the second decoupling
transition for the incommensurate system is the same as the sole decoupling
transition found in the commensurate system,
where all the particles in $s_1$ begin to slip with respect to the particles
in $p$.
The two step decoupling transition for the incommensurate system
appears due to the presence of vacancies in $s_1$.  At commensuration,
all of the particles in $s_1$ are located within potential minima created by
the spacing of the particles in $p$.  
Below commensuration, a fraction of the sites
in this periodic potential are empty, producing 
effective vacancies in $s_1$.
In the locked phase at $F_D<1.04$, all the particles in $s_1$ move 
at the same velocity as the particles in $p$.
At the first decoupling transition, the vacancies in $s_1$ begin to slip
with respect to the particles in $p$.  This can be viewed as
a depinning transition.  Every time a vacancy slips, only 
one of the particles in $s_1$ slips with respect to $p$ while the 
remaining particles in $s_1$ stay locked with $p$.
As a result,
most of the particles in $s_1$ continue to increase in 
velocity with increasing $F_{D}$.
For drives above the second decoupling transition, all of the particles 
in $s_1$ slip with respect to $p$ and the slipping is no longer dominated
by the motion of vacancies.

In Fig.~\ref{fig:4}(d) we plot $V_p$ and $V_{s1}^{\prime}$ versus $F_D$ for the 
same system in Fig.~\ref{fig:4}(c) but with $R_{s1,p} = 1.08$,  
where $N_{s1}>N_p$ so that a few incommensurate particles appear in $s_1$.
The overall shape of the velocity-force 
curve in this case is very similar to that for 
$R_{s1,p} = 0.92$ shown in Fig.~\ref{fig:4}(c), 
with a first decoupling occurring at a lower
drive of $F_{D} = 0.6$ than that for $R_{s1,p} = 0.92$, 
and a second decoupling transition 
occurring close to $F_{D} = 2.0$. 
Here the incommensurations in $s_1$ form doubly occupied sites in the 
periodic potential created by the particles in $p$.
At $R_{s1,p}=0.92$ when there are vacancies in $s_1$, slipping of
a particle adjacent to a vacancy occurs because the particle is able to move
closer to the barrier separating two minima in the periodic potential.
This is because
the force the particle experiences on one side from a
neighboring particle in $s_1$ is not compensated due to the missing particle
at the vacancy site.
As a result, there is an extra force of the order of $F^{pp}(a)$ 
on the slipping particle, where $a$ is the lattice constant
of the particles in $p$. 
For the doubly occupied sites at $R_{s1,p}=1.08$, a similar situation
occurs;
however, the slipping particle in $s_1$ is located at a doubly occupied site
and feels an uncompensated force from the other particle located within
the same site.  The extra force in this case is
$F_{pp}(a^{\prime})$, where $a^{\prime}<a$ 
in order for the site to be doubly occupied. 
Thus, $F_{pp}(a^{\prime})<F_{pp}(a)$, so the initial decoupling transition
occurs at a lower value of $F_D$ for samples with $R_{s1,p}>1$ that have
incommensurations than for samples with $R_{s1,p}<1$ that contain vacancies.

The appearance of multiple decoupling transitions 
just below and above commensuration and only 
one transition at commensuration
is similar to the single and multiple depinning
transitions observed in vortex systems \cite{Periodic} 
and colloidal systems \cite{Olson2} with periodic potentials at 
and near commensuration. In these 2D systems, at the matching 
filling of 1 particle per substrate minimum 
there is a single transition from a pinned state to a flowing state, 
while at fillings slightly away from commensuration,
well-defined vacancies or interstitial particles appear which
are highly mobile and depin at a lower external drive than 
the commensurate particles.   
The 2D systems are generally more complicated and allow for 
more than two depinning transitions near but not at 
commensuration \cite{Periodic}; 
however, as in our 1D case, it is the presence of 
two types of particles, the commensurate particles and the interstitial or 
vacancy sites, that produce the multiple depinning transitions. There are some
important differences between our two channel system and
the 2D vortex and colloidal systems.  Our system contains 
no fixed periodic substrate so there is no pinned phase; 
however, there is a moving fully coupled state  
which is analogous to the pinned state. 
The regime in which the vacancies or incommensurations slip 
is then analogous to the depinning transitions of interstitials or vacancies, 
and the high driving phases at which all the particles are slipping 
corresponds to the completely depinned regime in the 2D systems.  

\begin{figure}
\includegraphics[width=3.5in]{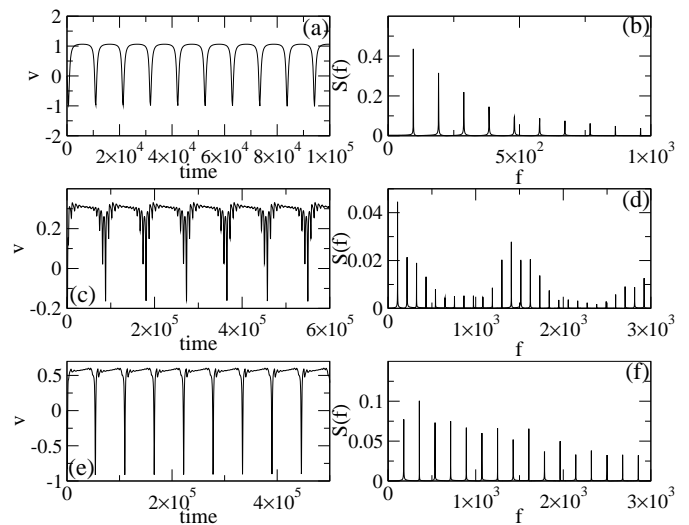}
\caption{  
(a) The velocity $v$ of a single particle in $s_1$ 
vs time for the commensurate system in Fig.~\ref{fig:2}(a) at $R_{s1,p} = 1.0$
and $F_{D} = 2.25$. (b) The Fourier transform $S(f)$ of the signal in
panel (a) shows a well-defined characteristic frequency.
(c) $v(t)$ for a system with
$R_{s1,p} = 1.08$ at $F_{D} = 0.65$.
(d) The corresponding $S(f)$ shows that there are two 
frequencies present.
(e) $v(t)$ for a system with $R_{s1,p} = 0.92$ at $F_D=1.1$. 
(f) The corresponding $S(f)$ shows the presence of
two frequencies.
}
\label{fig:5}
\end{figure}

In order to show more clearly that the particles in $s_1$ 
are experiencing a periodic potential produced by the particles
in $p$ and that there are two effective types of particles in
$s_1$ away from commensuration,
in Fig.~\ref{fig:5}(a) we plot the 
time trace of the velocity $v(t)$ of a single particle 
in $s_1$ for the system in Fig.~\ref{fig:2}(a) at $R_{s1,p} = 1.0$ 
in the locked phase at $F_{D} = 2.25$. 
The value of $v(t)$ is nearly constant except during the periodic
slip events, during which $v$ drops briefly below zero indicating that the
particle temporarily moves backwards.
In Fig.~\ref{fig:5}(b) we show the Fourier transform $S(f)$ of $v(t)$
highlighting the presence of a single characteristic frequency determined by
the slipping events.
In the locked phase, 
there is no high-frequency oscillation of the velocity of any of the particles. 
In Fig.~\ref{fig:5}(c) we plot $v(t)$ 
for the system with $R_{s1,p} = 1.08$ 
from Fig.~\ref{fig:4}(d) at $F_{D} = 0.65$
where the channels are not completely locked 
but where $V_{s1}^{\prime}$ is still increasing with increasing $F_{D}$, and
in Fig.~\ref{fig:5}(d) we show $S(f)$ for the same data. 
There are now two frequencies present.  The lower frequency is produced
by the same slipping events that occurred for the commensurate system in
Fig.~\ref{fig:5}(a,b), while the higher frequency originates from the motion of
the incommensuration through $s_1$.
In Fig.~\ref{fig:5}(e,f) we plot $v(t)$ and $S(f)$ for 
a sample with $R_{s1,p} = 0.92$ at $F_{D} = 1.1$.  Again, there are
two characteristic frequencies. The lower frequency is associated with the
same slipping events shown previously, while the higher frequency is produced
by the motion of a vacancy through $s_1$, rather than by the motion of an
incommensuration.

\begin{figure}
\includegraphics[width=3.5in]{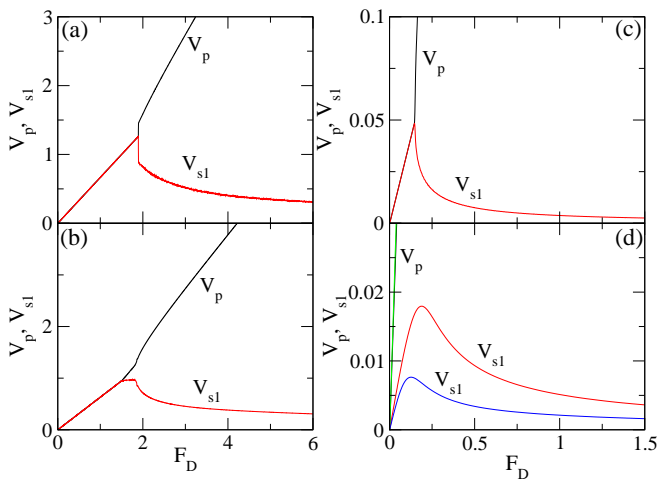}
\caption{ 
$V_{p}$ (upper curve) and $V_{s1}$ (lower curve)
vs $F_D$ for the two channel system from Fig.~\ref{fig:2}(a). 
(a) At $R_{s1,p} = 0.5$, there is a single transition from the 
completely coupled state to the partially decoupled state. 
(b) At $R_{s1,p} = 0.58$ there are two transitions. 
The first decoupling transition occurs when the incommensurations begin
to slip while the other particles in $s_1$ remain locked with $p$.
(c) At $R_{s1,p} = 2.0$ there is a single transition out of the locked phase.
(d) At 
$R_{s1,p}=1.92$ (upper $V_{s1}$ curve) and 
$R_{s1,p} =2.08$ (lower $V_{s1}$ curve) there is no locked phase, but
there is still a local maximum in $V_{s1}$ 
which is higher for 
$R_{s1,p} = 1.92$. 
}
\label{fig:6A}
\label{fig:6B}
\end{figure}

The two stage decoupling process is most pronounced 
for fillings close to $R_{s1,p} = 1.0$, but the same effects appear
near certain fractional ratios. 
For example, in Fig.~\ref{fig:6A}(a) we plot $V_p$ and $V_{s1}$ versus $F_D$ for 
a sample with $R_{s1,p} = 0.5$ which has a single sharp decoupling 
transition at $F_{D} = 1.9$.  Just above this filling at
$R_{s1,p}=0.58$, shown in Fig.~\ref{fig:6A}(b),
there is an initial decoupling transition 
of the incommensurations near $F_{D} = 1.5$ into a state where 
$V_{s1}$ still increases with increasing $F_D$ but with a greatly
reduced slope. 
A second decoupling transition appears at $F_{D} = 1.8$, 
and above this drive $V_{s1}$ decreases with increasing $F_{D}$.
In this case, incommensurations appear with respect to the particle
configuration that occurs at $R_{s1,p}=0.5$.  There is a single frequency
associated with the motion of the particles in $s_1$ for $R_{s1,p}=0.5$,
while for the incommensurate case of $R_{s1,p}=0.58$, two frequencies are
present.
This trend persists for higher values of 
$R_{s1,p}$ as shown in Fig.~\ref{fig:6B}(c) at $R_{s1,p} = 2.0$. Here there
is a single sharp decoupling transition,
while just below and just above this filling at
$R_{s1,p} = 1.92$ and $R_{s1,p}=2.08$, Fig.~\ref{fig:6B}(d) shows that
the locking phase is absent but that a 
strong local maximum in $V_{s1}$ appears
at $F_{D} = 0.19$ for $R_{s1,p} = 1.92$ 
and at $F_{D} = 0.125$ for $R_{s1,p} = 2.08$. 

\begin{figure}
\includegraphics[width=3.5in]{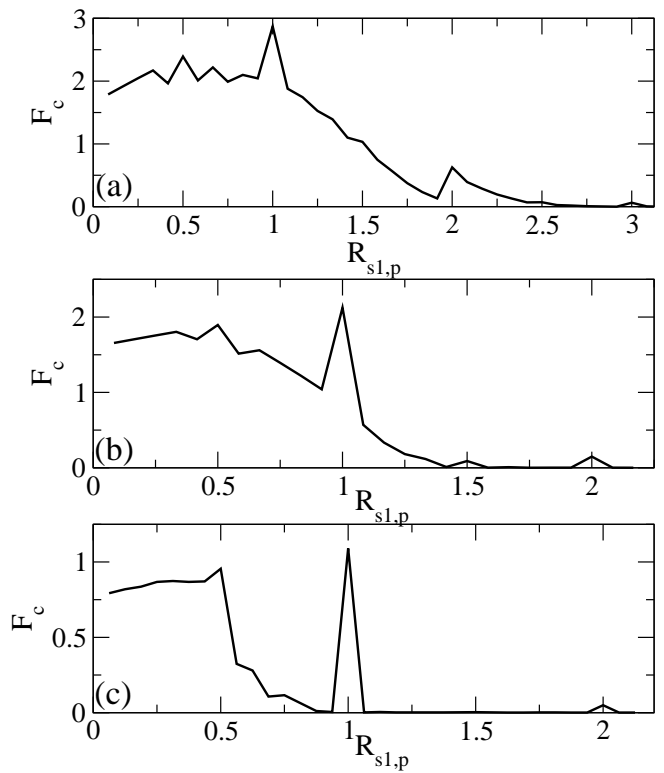}
\caption{ 
The force $F_{c}$ at the transition from the locked to unlocked phase 
vs $R_{s1,p}$ in two channel samples with different values of $a$.
(a) 
At $d/a=0.44$, commensuration peaks occur at 
$R_{s1,p} = 1.0$, 2.0, and $3.0$.
Fractional peaks and anomalies 
appear at $R_{s1,p} = 0.5$, 1.5, and $2.5$.
(b) At 
$d/a=0.67$, there are commensuration peaks 
at $R_{s1,p} = 0.5$, 1.0, 1.5, and $2.0$. 
(c) At $d/a = 1.0$, 
the strongest commensuration peaks appear at $R_{s1,p}=1$ and 0.5.
}
\label{fig:7}
\end{figure}

By performing a series of simulations for 
varied $R_{s1,p}$, we determine the location $F_c$ of the transition from
complete locking to a decoupled state
and map out where the commensuration effects occur. 
In Fig.~\ref{fig:7}(a), we plot the decoupling force $F_{c}$ versus
$R_{s1,p}$ for a two channel sample with 
$d/a = 0.44$, which falls in the strong coupling regime in Fig.~\ref{fig:3}(b).
There are peaks in $F_c$ at $R_{s1,p} = 1.0$, 2.0, and $3.0$, 
along with submatching peaks at $R_{s1,p} = 1/3$, $1/2$, and $2/3$.
Additionally, weaker anomalies appear at 
$R_{s1,p}=1.5$ and $2.5$. 
In Fig.~\ref{fig:7}(b) we show $F_c$ versus $R_{s1,p}$ for a sample with
$d/a = 0.67$.  The value of $F_c$ at 
the commensurate filling of $R_{s1,p}=1.0$ is lower for the $d/a=0.67$ sample
than for the $d/a=0.44$ sample.
Figure \ref{fig:7}(b) also has
clear peaks in $F_c$ at $R_{s1,p} = 2.0$, 1.5, and $0.5$, 
while above $R_{s1,p}=2.0$ 
within our resolution there are no peaks or regions where
the system is locked.
In the regions with $F_c=0$ where the locked phase is absent,
the second decoupling transition still appears at higher drives and
can be detected as the point at which
$V_{s1}$ changes from increasing to decreasing with increasing $F_{D}$.
For higher particle densities and fixed $d$, 
the commensurability effects still persist as shown in Fig.~\ref{fig:7}(c)
for $d/a = 1.0$. 
Here, peaks in $F_c$ occur at $R_{s1,p} = 0.5$, 1.0, and $2.0$.  

The appearance of the commensuration effects 
at integer and fractional fillings suggests that this system
exhibits the same behavior 
found for the depinning of repulsively 
interacting particles on a 1D
fixed periodic potential; 
however, there are several differences between the two systems.
For particles on a fixed periodic potential, 
the depinning force $F_{c}$ at fields where the 
particle-particle interactions
cancel due to symmetry equals the maximum value of the pinning force 
$F_{p}$ so that  $F_{c}  = F_{p}$ at fillings 
$1/12$, 1/8, 1/6, 1/4, 1/2, and $1.0$.
For the drag system shown in Fig.~\ref{fig:7}, this does not occur and
there is even a trend for $F_{c}$ to increase 
at the lowest fillings. 
This is because the 
substrate potential created by the particles in $p$ is 
not fixed but can distort since the particles
in either channel can shift. 
At $R_{s1,p} = 1.0$, 
the periodic potential is fairly rigid due to the matching of the particle
positions in $p$ and $s1$, and
any distortion of the particles in $p$ is energetically unfavorable.
In contrast,
at very low fillings such as 
$R_{s1,p}=0.125$, the particles in $p$ distort near the locations of the
particles in $s_1$ in order to create a localized lowering of the density
in $p$ above each particle in $s_1$.
As a result, the particles in $s_1$ no longer experience the same periodic
potential from $p$ that was present for the commensurate case
of $R_{s1,p}=1.0$.
Even at $R_{s1,p} = 0.5$, the particles in $p$ can distort, reducing
the strength of the coupling to the particles in $s_1$.

\begin{figure}
\includegraphics[width=3.5in]{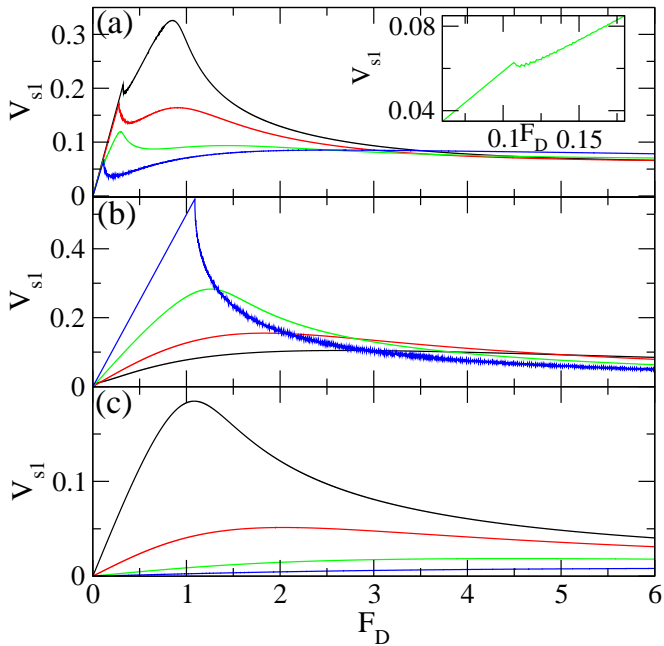}
\caption{ 
$V_{s1}$ vs $F_{D}$ for the
system in Fig.~\ref{fig:7}(b) with $d/a=0.67$. 
(a) $R_{s1,p} = 0.562$, 0.625, 0.6875, and $0.75$, from top to bottom. 
Inset: Detail of the $R_{s1,p}=0.6875$ curve from the main panel.
(b) $R_{s1,p} = 0.8125$, 0.875, 0.9375, and $1.0$, from bottom to top.
(c) $R_{s1,p} = 1.0625$, 1.125, 1.1875, and $1.25$, from top to bottom.  
}
\label{fig:8}
\end{figure}

In order to better understand the changes 
in dynamics at the different fillings, in Fig.~\ref{fig:8}
we plot $V_{s1}$ 
as a function of $F_D$                   
for varied $R_{s1,p}$ in a system with $d/a=0.67$.
At $R_{s1,p}=0.5$, a single decoupling transition occurs and
$V_{s1}$ is a monotonically decreasing function.
For $R_{s1,p} = 0.562$ and $0.625$, shown in Fig.~\ref{fig:8}(a), 
there is a clear double peak structure in $V_{s1}$ with
one peak falling at the depinning of the incommensurations and 
the second peak appearing at the unlocking transition.  
At $R_{s1,p} = 0.6875$ in Fig.~\ref{fig:8}(a), 
there is now a three peak structure
in $V_{s1}$.  The first peak, shown in the inset of 
Fig.~\ref{fig:8}(a),
falls at the transition out of the completely locked phase at $F_{D} = 0.11$. 
The second and largest peak is at $F_{D} = 0.3$, 
while a third broad peak also appears that is centered 
at $F_{D} = 1.45$. The broad peak is the remnant of 
the second peak in $V_{s1}$ found
for $R_{s1,p} = 0.562$ and $0.625$; with increasing $R_{s1,p}$, this peak
broadens and the center shifts to higher values of $F_D$.
For $0.11 < F_{D} < 0.3$, the particles in $s_1$ are almost completely
locked but there is a single incommensuration which has begun to slip.
For $R_{s1,p} = 0.75$, the initial peak is lost and the decoupling transition
peak now falls at $F_D=0.11$. 
There is also a very broad maximum centered at $F_{D} = 4.0$.
Another interesting feature is that at higher 
$F_{D}$ such as at $F_{D} = 6.0$, $V_{s1}$ for
$R_{s1,p} = 0.75$ is higher than $V_{s1}$ 
at the lower values of $R_{s1,p}$,  
even though at low $F_{D}$ $R_{s1,p}$ showed the lowest value of 
$V_{s1}$.
This suggests that at high values of $F_{D}$, additional
drag is produced by the interaction between the incommensurations in $s_1$
and the particles in $p$.

In Fig.~\ref{fig:8}(b) we plot $V_{s1}$ versus $F_D$ 
for $R_{s1,p} = 0.8125$, 0.875, 0.9375, and 1.0.
The maximum value of $V_{s1}$ 
increases as $R_{s1,p}$ increases toward $R_{s1,p} = 1.0$ 
and the broad maximum in $V_{s1}$
sharpens and shifts toward lower $F_{D}$. 
Here, $R_{s1,p}=1.0$ has the lowest value and $R_{s1,p}=0.8125$ has the 
highest value
of $V_{s1}$ 
at $F_{D} = 6.0$.
In Fig.~\ref{fig:8}(c) we show $V_{s1}$ versus $F_D$ for 
$R_{s1,p} = 1.0625$, 1.125, 1.1875, and $1.25$.
For these values of $R_{s1,p}$ there is no completely locked phase; 
however, there is still a peak feature in $V_{s1}$
for $R_{s1,p}= 1.0625$ and $1.125$ which broadens and shifts to higher $F_D$ for
increasing $R_{s1,p}$.
At $R_{s1,p} = 2.0$ a locked phase reappears and the shape of 
$V_{s1}$ versus $F_{D}$ 
is very similar to the curve shown for $R_{s1,p} = 1.0$.    

\begin{figure}
\includegraphics[width=3.5in]{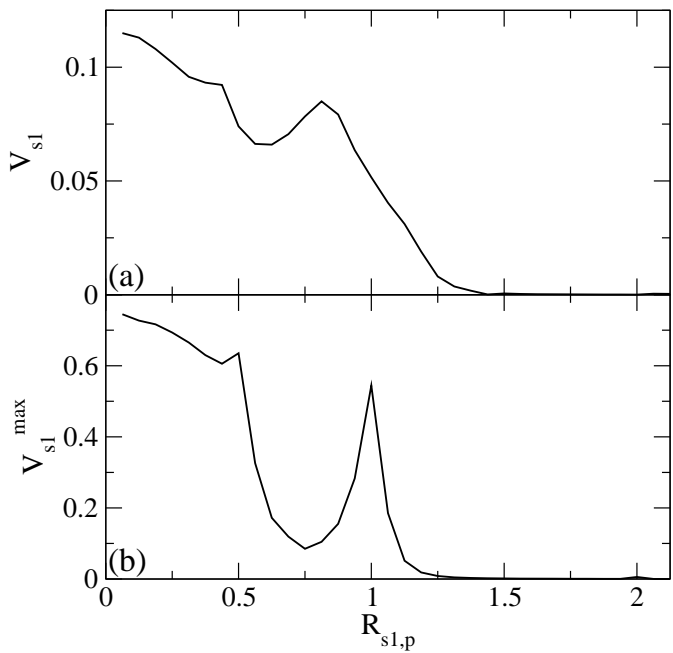}
\caption{ 
$V_{s1}$ vs $R_{s1,p}$ 
at $F_{D} = 6.0$ for the two channel system 
with $d/a=0.67$.  Here the 
peaks appear not at the commensurate fields 
of $R_{s1,p} = 1.0$ and $R_{s1,p} = 0.5$ but at 
$R_{s1,p} = 0.8$ and $R_{s1,p}=0.45$. 
(b) $V_{s1}^{max}$ vs $R_{s1,p}$ 
obtained for each filling at the $F_{D}$ 
where $V_{s1}$ reaches its maximum value.  The
peak centered at $R_{s1,p} = 1.0$ is much broader than 
the peak in $F_c$ at $R_{s1,p}=1.0$ shown in Fig.~\ref{fig:7}(b).  
}
\label{fig:9}
\end{figure}

In Fig.~\ref{fig:9}(a) we plot $V_{s1}$ 
versus $R_{s1,p}$ at a fixed drive of $F_{D} = 6.0$ for the
system in Fig.~\ref{fig:8} with $d/a=0.67$.
Larger values of 
$V_{s1}$ indicate 
that the particles in $s_1$ are exerting a larger drag on the
particles in $p$.
Here, peaks fall at $R_{s1,p}  = 0.45$ and $R_{s1,p} = 0.8$, rather than
at the values $R_{s1,p}=0.5$ and $R_{s1,p}=1.0$ where peaks appeared in 
$F_c$ in Fig.~\ref{fig:7}(b).  
This shows that at high drives, the drag by the $s_1$ particles is the 
most effective away from the commensurate fillings.
The curves shown in Fig.~\ref{fig:8} indicate that if 
$V_{s1}$ were
measured at a lower value of $F_{D}$, the peaks in Fig.~\ref{fig:9}(a)
would shift closer to $R_{s1,p} = 0.5$ and $1.0$.

In vortex systems with 2D periodic pinning arrays, 
experiments have shown that the pinning is enhanced at the matching fields 
as indicated by dips in the resistivity for low applied drives.
For vortices that are strongly driven, however,
the resistivity dips were found to shift away from 
the integer matching fields \cite{Jiang}. 
The interpretation was 
that at the matching fields in the highly driven system,
the vortices form a very ordered moving commensurate state, 
while at the incommensurate fields the moving state is not as well ordered
and thus the effectiveness of the pinning increases away from commensuration
at high drives.
Although the disorder in the incommensurate state 
causes the system to begin slipping at a lower drive
for incommensurate fields,
at high drives the disordered state 
experiences more fluctuations than the ordered state 
which induce some additional
drag.

In Fig.~\ref{fig:9}(b) we plot $V_{s1}^{max}$ versus $R_{s1,p}$. 
Here the measurement of $V_{s1}^{max}$ is performed
not at a fixed $F_{D}$ but at the $F_{D}$ where 
$V_{s1}$ reaches its maximum for each value of $R_{s1,p}$. 
In this case, a strong peak in $V_{s1}^{max}$ 
appears at $R_{s1,p} = 1.0$. This peak is wider than the peak in
$F_c$ at $R_{s1,p}=1.0$ in Fig.~\ref{fig:7}(b) due to the fact that
the maximum value of $V_{s1}$ increases as 
$R_{s1,p} = 1.0$ is approached, as shown in Fig.~\ref{fig:8}(b,c).  

\begin{figure}
\includegraphics[width=3.5in]{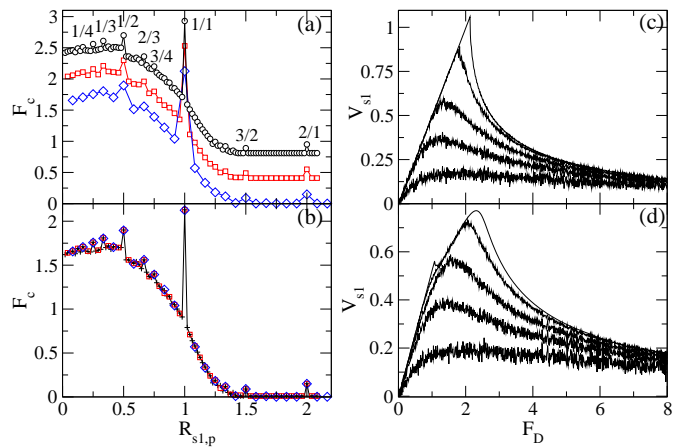}
\caption{ 
(a) $F_{c}$ vs $R_{s1,p}$ for the system with $d/a = 0.67$ from 
Fig.~\ref{fig:7}(b) of length $L$ (diamonds), $2L$
(squares, curve shifted up by 0.4), and $4L$
(circles, curve shifted up by 0.8). 
For the larger systems, higher order fractional
peaks appear at $R_{s1,p} = 1/4$, 1/3, 1/2, 2/3, 3/4, and $3/2$.  
(b) The same data plotted without vertical shifts, for system sizes
of $L$ (diamonds), $2L$ (squares), and $4L$ (plus signs).  Connecting
lines are drawn only for the $4L$ system.  The curves
overlap exactly and the values of $F_c$ are unaffected by system size.
(c,d) $V_{s1}$ vs $F_{D}$ for the $4L$ system from (a) at (c) $R_{s1,p} = 1.0$
and (d) $R_{s1,p} = 0.896$ for temperatures of
$T = 0.0$, 0.22, 0.88, 2.0, and $4.5$, from top to bottom. 
The decoupling transition drops to lower $F_D$ with increasing $T$, while
the drag effects persist up to high temperatures.
}
\label{fig:new}
\end{figure} 

\subsection{Finite Size and Temperature Effects}

To determine whether further higher order submatching effects in 
$R_{s1,p}$ can be resolved for larger systems and whether 
the values of $F_{D}$ at which the unlocking transitions occur change with
system size,
we consider the system at $d/a = 0.67$ from Fig.~\ref{fig:7}(b) and
analyze $F_{c}$ for samples of size $2L$ and $4L$.  Here we hold $a$ fixed by
increasing $N_p$ to 32 and 64, respectively.
In Fig.~\ref{fig:new}(a) we plot $F_{c}$ versus $R_{s1,p}$ for samples of size
$L$, $2L$, and $4L$, with the curves shifted vertically for clarity.
In the larger samples, there are clearly fractional peaks falling at
$R_{s1,p} = 1/4$, 1/3, 1/2, 2/3, 3/4, and $3/2$.
The $4L$ sample even shows some evidence of a peak at $R_{s1,p}= 1/8$.
We expect that for even larger systems, even more fractional peaks will 
appear but that the higher order peaks will be increasingly weak in size,
similar to the behavior of fractional peaks observed in other systems such as
vortices on periodic substrates \cite{Baert,Nori}.
In Fig.~\ref{fig:new}(b) we plot the same data without vertical shifts to show
that the depinning thresholds for the three systems
overlap exactly; only the resolution is changed by the system size. 
We find no changes in the velocity-force curves as the size of the
sample is increased,
indicating that the system sizes we are studying capture the
essential behavior. 
We also find a similar lack of dependence on sample size
for the three layer systems that are described in Section IV.
We note that for commensurate-incommensurate systems such as 
the Frenkel-Kontrova model \cite{Bak}, submatching effects theoretically
occur for all rational values of $m/n$, where $m$ is the number of particles
and $n$ is the number of substrate minima.
In the Frenkel-Kontrova model, true incommensurate behavior occurs only
for systems of infinite size at irrational filling ratios. 
In our system the higher order submatching effects are destroyed due to the
fact that we do not have a fixed substrate; instead, the effective substrate
experienced by the particles in one channel due to the presence of particles
in the neighboring channel is able to distort.  As a result, our system does
not map directly onto commensurate-incommensurate systems such as the
Frenkel-Kontrova model, although it displays several similarities with such
models as we have shown.

In an experimental realization of the system we propose, such as with
colloids confined to channels, thermal effects will be present.  To test
the stability of the different regimes we observe against thermal perturbations,
we have performed simulations with the $4L$ system with $d/a=0.67$
at $R_{s1,p} = 1.0$ and $R_{s1,p} = 0.896$ at
finite temperature. 
We use the same procedure employed in previous works to model 
the thermal fluctuations \cite{Libal2}.  We add
a Langevin noise term $F^{T}_{i}$ to the equation of motion 
with the properties
$\langle F^{T}_{i}(t)\rangle = 0$ and 
$\langle F^{T}_{i}(t)F^{T}_{j}(t^\prime)\rangle = 2\eta k_{B} T \delta_{ij}\delta(t-t^\prime)$. 
Fig.~\ref{fig:new}(c) illustrates $V_{s1}$ versus $F_D$
at $R_{s1,p} = 1.0$ and 
Fig.~\ref{fig:new}(d) shows 
$V_{s1}$ versus $F_D$ at $R_{s1,p} = 0.896$ 
for $T = 0.0$, 0.22, 0.88, 2.0,
 and $4.5$. 
As the temperature increases, the value of $F_{c}$ decreases until
for $T > 2.0$ the locked phase has almost completely vanished; 
however, drag effects on the secondary channel continue
to persist up to much higher temperatures. 
For $R_{s1,p} = 0.896$, the locking phase is lost at lower $T$ than
for $R_{s1,p}=1.0$ due to the fact that the effective incommensurations 
are more mobile and hence require a smaller level of thermal fluctuations to
escape from the potential minima.
These results show that the drag and locking features described for the
zero temperature system should persist
under finite temperature 
provided that the thermal fluctuations are not excessively strong.    

Possible experimental realizations of the two channel system include modified
versions of the colloidal experiments which have already been performed on
coupled one-dimensional channels \cite{Bleil}.
Colloidal systems are subject to thermal fluctuations and hydrodynamic
interactions which can arise in the surrounding fluid.  We showed above that 
the dynamic phases are robust against
moderate thermal fluctuations.  There is ongoing discussion regarding how
the inclusion of hydrodynamic effects would impact the dynamics of driven
colloidal systems.  Recent two-dimensional simulations of an electrophoretically
driven charged colloidal system similar to the one we consider showed that
when the charge on the colloids are sufficiently strong, the dynamical 
behavior of the system is not altered by the addition of hydrodynamic
interactions \cite{LowenN}.
Due to the good agreement that has been found between numerous simulations
of driven colloid systems in which hydrodynamic effects are neglected and the
actual behavior of driven colloids in experiment, we expect that at least some
of the features that we describe should be observable in a colloidal realization
of this system.
Another possible realization would be to generalize the recent experiments
performed with dusty plasmas interacting with a one-dimensional groove to
create what are termed Yukawa chains \cite{Goree}.  If more than one groove
were created in the substrate, it should be possible to couple two or more
of the Yukawa chains, to use a laser to drive one of the chains, and then
to analyze the response of the secondary chain.
In this case inertial effects could modify the behavior
since dusty plasma systems are generally not in the
overdamped limit.    

\subsection{Ratchet Effect With ac Drives}

We next show that when the particles in $p$ are driven with an ac drive,
it is possible 
to generate a net dc motion of the particles in $s_1$
or a ratchet effect. Ratchet effects 
produced by applied ac drives have
been studied extensively in systems of particles interacting 
with asymmetric substrates \cite{Reimann}; however, it is 
also possible to create a 
ratchet effect in the absence of an asymmetric substrate
when the ac drive has certain asymmetries and when the response
of the system is nonlinear
\cite{Sav,Hastings,Cole,Caplin}. 
This type  of ratchet has been realized in
systems with two interacting species of 
superconducting vortices such as when 
Josephson vortices couple to pancake vortices \cite{Sav,Cole,Caplin}, as
well as in interacting binary colloidal systems where only
one colloid species couples to an external driving field 
and produces a rectification of the 
other colloid species \cite{Hastings}. 

\begin{figure}
\makebox[0pt][l]{
\includegraphics[width=3.5in]{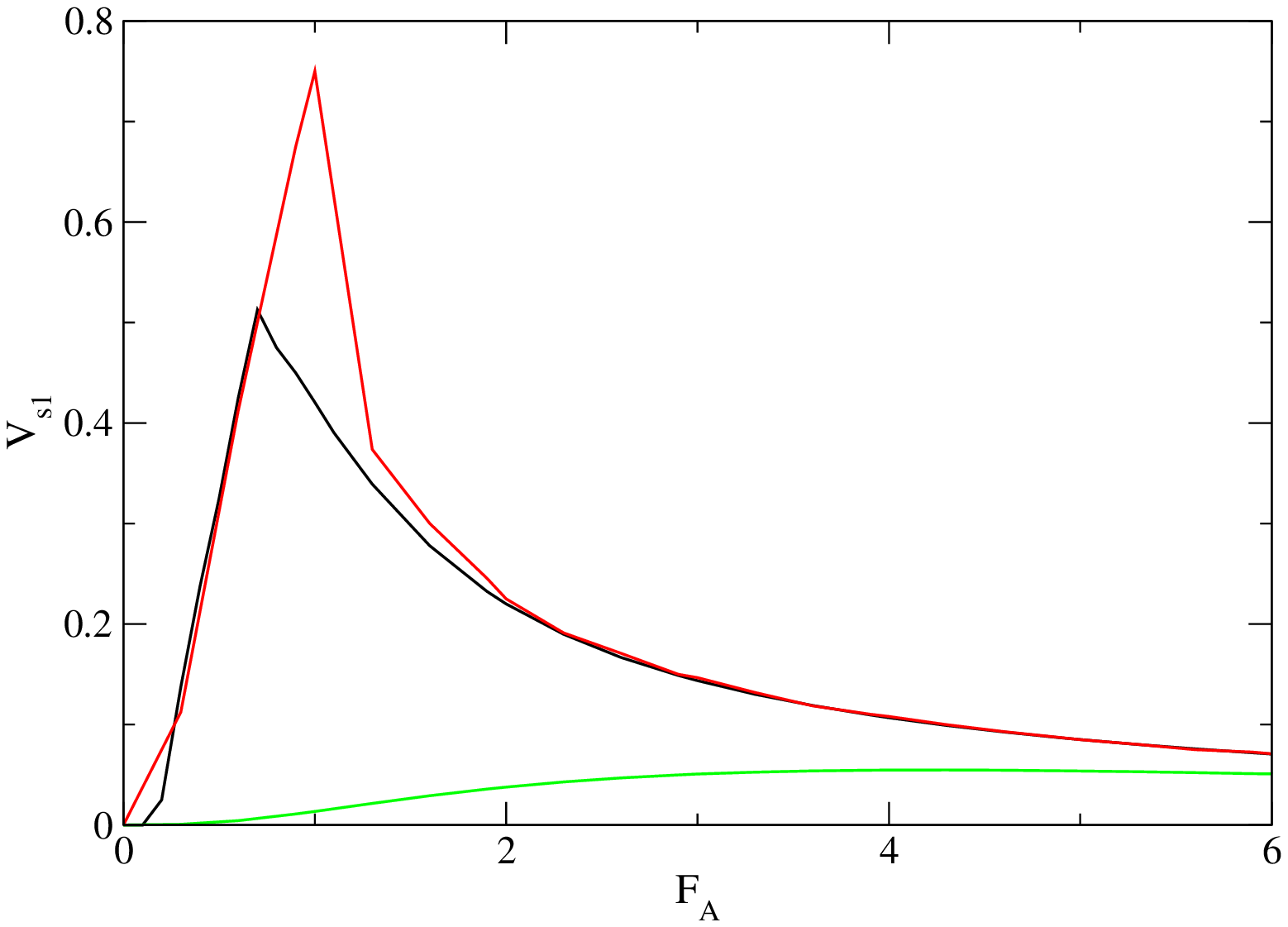}}%
\includegraphics[width=3.5in]{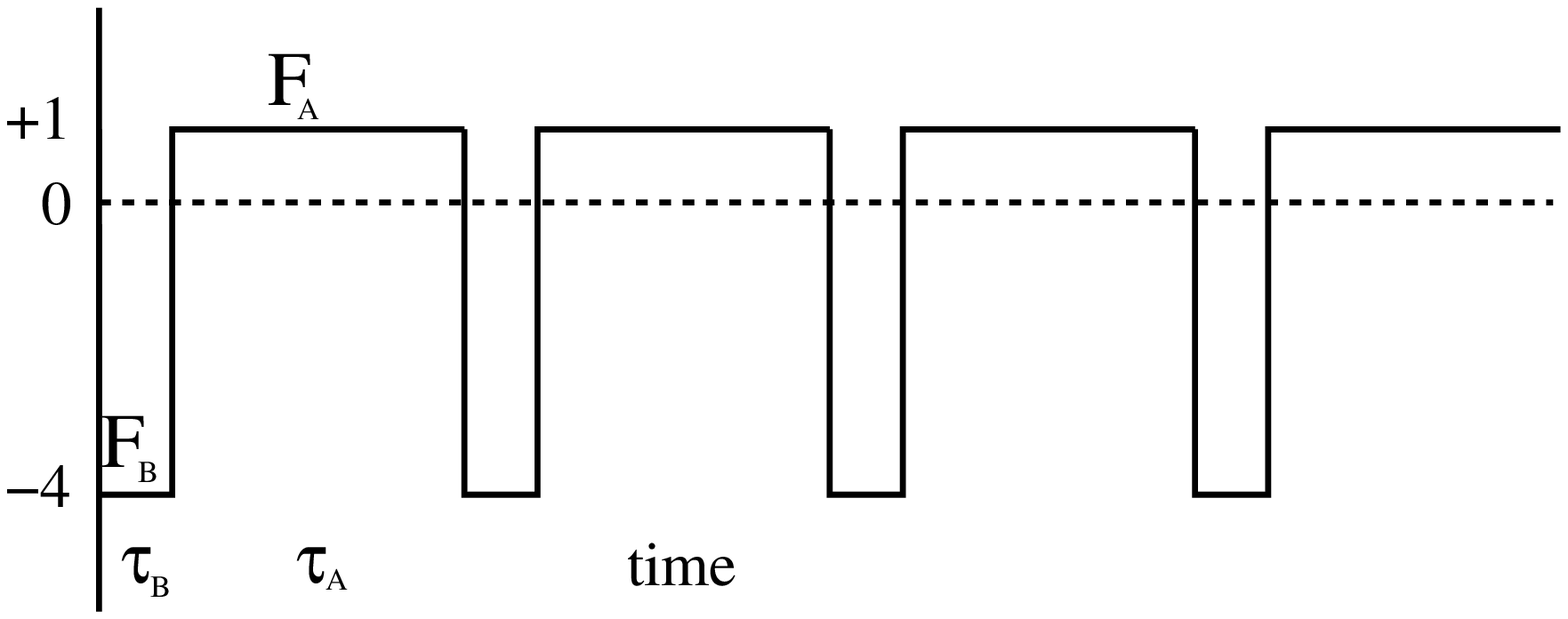}
\caption{ 
Inset: Schematic of the ac force applied to $p$.
Each period $\tau$ is divided into two parts. 
A force $F_{A}$ is applied in the
positive direction for a duration $\tau_{A}$, then
a force $F_{B}$ is applied in the negative direction
for a duration $\tau_{B}$, 
with the condition that $F_{B}/F_{A} = \tau_{A}/\tau_{B} = 4.0$. 
Main panel:
Induced dc velocity $V_{s1}$ averaged over
multiple ac drive periods vs $F_{A}$ 
for a two channel system with $d/a = 0.67$ under the applied ac drive
shown schematically in 
the inset.
Upper curve: $R_{s1,p}=1.0$; middle curve: $R_{s1,p}=0.75$; lower curve:
$R_{s1,p}=1.25$.
Here the ratchet effect reaches its maximum value
for $R_{s1,p} = 1.0$.
Within this range of $F_A$, $V_{p} = 0$ when averaged over an ac drive 
period.      
}
\label{fig:10}
\label{fig:11}
\end{figure}

Here we consider an ac square drive applied only to $p$.  The period $\tau$
of the square drive is divided unevenly into two parts as illustrated
in the inset of Fig.~\ref{fig:10}.
In part $A$, we apply a force ${\bf F}_D=F_A{\bf \hat x}$ 
in the positive direction
for a duration $\tau_{A}$, while in part $B$ we apply a force 
${\bf F}_D=-F_B{\bf \hat x}$ in
the negative direction for a duration $\tau_B=\tau-\tau_A$.  In selecting
$F_A$ and $F_B$, we impose the condition
$F_A\tau_A - F_B\tau_B=0$ so that there is no net dc drive.
If the response of the system is perfectly linear, this drive will not
generate a net dc motion of the particles in either channel.
On the other hand, if the coupling between $s_1$ and $p$ is nonlinear,
it is possible to induce a dc motion of the particles in $s_1$ by
applying this ac drive to the particles in $p$.
If both $F_{A}$ and $F_{B}$ 
are below the first decoupling transition $F_c$, the motion of the particles
in both channels is completely locked, the response is perfectly linear,
and there is no ratchet effect.
If $F_{A} < F_{c}$ and $F_{B} > F_c$, 
a net dc drift of the particles in $s_1$ will occur
since the particles in $s_1$ remain completely locked with the particles
in $p$ during part $A$ of the drive cycle, but during part $B$ of the 
cycle the particles in $s_1$ are partially decoupled and do not move
all the way back to their starting position by the end of the cycle.

In Fig.~\ref{fig:11} we illustrate the ratchet effect which produces a finite
positive value of $V_{s1}$ under the ac drive described
above.  We fix $F_B/F_A=4.0$ and $d/a=0.67$, and plot
the time-averaged $V_{s1}$ versus 
$F_{A}$ for  $R_{s1,p} = 0.75$, 1.0, and $1.25$.
For low $F_{A}$, $V_{s1}$ starts small but rapidly grows 
with increasing $F_A$, reaching a sharp peak for 
$R_{s1,p} = 0.75$ and $R_{s1,p}=1.0$.  
As $F_A$ increases above this peak,
$V_{s1}$ gradually decreases with increasing $F_A$ since 
the drag effect becomes smaller for higher drives 
as shown in Fig.~\ref{fig:2}(a). 
For $R_{s1,p} = 1.25$, the ratchet effect is strongly reduced but still
persists, 
indicating that the ratchet effect should be 
a robust feature for all fillings. 
In all cases there is no induced dc flow of the particles in $p$.
Our system can be regarded as containing 
two species of particles: the directly driven particles in $p$, and the 
undriven particles in $s_1$ that experience a drag from the particles in $p$.
It would be very interesting to 
look for a similar ratchet effect in coupled 
quantum wires in the regime where
Wigner crystallization may be occurring. 
This could be achieved 
by applying an ac drive of the type illustrated in 
the inset of Fig.~\ref{fig:10} to 
one wire and determining whether a dc response 
is induced in the second wire. 

\begin{figure}
\includegraphics[width=3.5in]{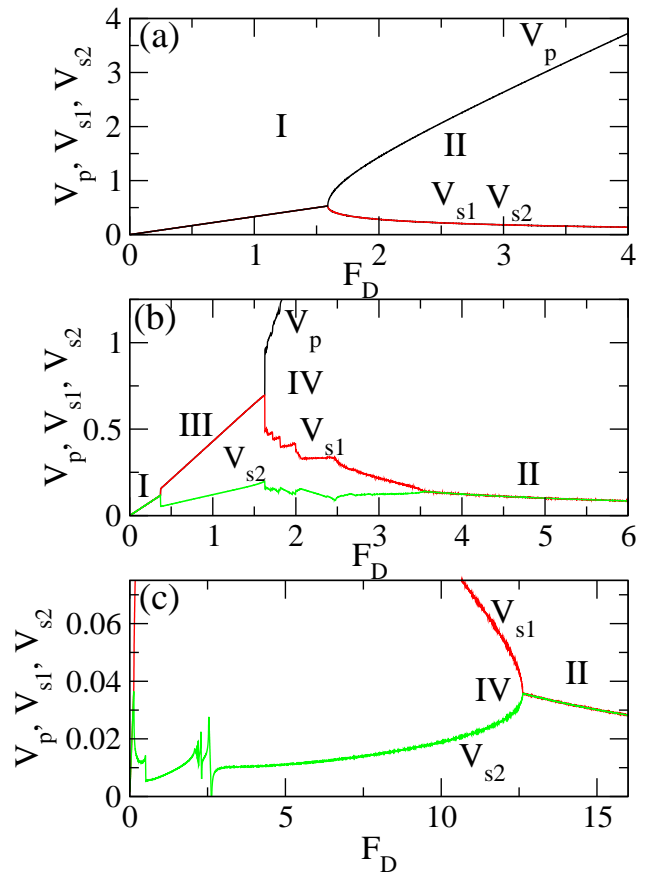}
\caption{ 
The velocities 
$V_{p}$, $V_{s1}$, and $V_{s2}$ vs $F_{D}$ for 
a three channel system
with $d/a=0.67$, $R_{s1,p} = 1.0$, and varied $R_{s2,s1}$. 
(a) At $R_{s2,s1} = 1.0$ there is a single transition from 
the locked region I to region II
where the particles in $s_1$ and $s_2$ remain locked with each other but
partially decouple from the particles in $p$.
(b) At $R_{s2,s1} = 1.16$ there 
is a transition from the locked region I to region III where
the particles in $s1$ and $p$ lock together but the particles in
$s_2$ partially decouple.  In region IV all three channels 
are unlocked, while at high $F_D$ the system enters region II when
the particles in $s_1$ lock with the particles in $s_2$ 
but are partially decoupled from the particles in $p$.
(c) At $R_{s2,s1} = 1.5$ the transition between regions IV and II
occurs at a much higher value of $F_D$.
}
\label{fig:12}
\end{figure}

\section{Three Channel Systems}
\subsection{Coupling-Decoupling Transitions for Partial Commensuration}

We next consider 
a system with three channels of particles where only the top channel is
subjected to a driving force.
We measure the velocities in each channel, denoted by 
$V_{p}$, $V_{s1}$, and $V_{s2}$, 
for particle ratios 
of $R_{s1,p} = N_{s1}/N_{p}$, $R_{s2,p} = N_{s2}/N_{p}$, 
and $R_{s2,s1} = N_{s2}/N_{s1}$.
For the commensurate case when all channels contain the same
number of particles, $R_{s1,p} = R_{s2,p}= R_{s2,s1} = 1.0$, 
the behavior is the same as in the two channel case at commensuration.
There is a single decoupling transition from 
region I, the completely locked phase, to region II,
where the particles in $s_1$ and $s_2$ remain locked
with each other but are partially decoupled from the 
particles in $p$.
This is illustrated in the plot of $V_p$, $V_{s1}$, and
$V_{s2}$ versus $F_D$ in Fig.~\ref{fig:12}(a) for a sample with
$d/a = 0.67$, $R_{s1,p}=1.0$, and $R_{s2,s1}=1.0$. 
The decoupling between the primary and the
secondary channels occurs at $F_{D} = 1.75$. 
The value of $F_{D}$ at decoupling is lower than for a sample containing
only two channels since the primary channel must now drag twice as many
secondary particles.

In Fig.~\ref{fig:12}(b) 
we plot the channel velocities versus $F_D$ for a sample with
$R_{s1,p}=1.0$ but with more particles in $s_2$, 
$R_{s2,s1} = 1.16$.
For $F_{D} < 0.37$ the system is in the completely locked region I,
while for $0.37 \leq F_{D} < 1.625$ the particles in $p$ and $s_1$ remain
locked but the particles in $s_2$ partially decouple.  We
term this range of $F_D$ region III, and in this region 
$V_{s2}$ still increases with increasing $F_{D}$.
For $1.625 \leq F_{D} < 3.56$, all three of the channels are unlocked; we
call this region IV. 
Within region IV, the velocity curves contain numerous small steps
associated with the intermittent coupling of the particles in $s_1$  
and $s_2$.
At the low $F_D$ end of region IV, $V_{s1}$ 
and $V_{s2}$ both decrease with increasing $F_D$,
but for $2.7 < F_{D} < 3.56$, $V_{s2}$ begins
to increase with increasing $F_D$ until $V_{s1}$ and
$V_{s2}$ join at the recoupling transition into region II.
Once the system is in region II, both $V_{s1}$ and $V_{s2}$
decrease monotonically with increasing $F_D$.

For samples with $R_{s1,p}= 1.0$ but 
with increasing $R_{s2,s1}$, the general features of the
velocity force curves are the same as Fig.~\ref{fig:12}(b), but the transition 
into region II is pushed to higher $F_{D}$.  This is illustrated
in Fig.~\ref{fig:12}(c) where we plot $V_p$, $V_{s1}$,
and $V_{s2}$ versus $F_D$ for 
a system with $R_{s1,p}=1.0$ and $R_{s2,s1} = 1.5$. 
Here region II does not appear until $F_{D} = 12.5$. 
Fig.~\ref{fig:12}(c) also shows more clearly 
the increase in $V_{s2}$ just below the onset of region II. 
For samples with $R_{s1,p} = 1.0$ and $R_{s2,s1} < 1.0$, 
only regions I and II occur and the velocity force curves
have the same form as the curves illustrated in Fig.~\ref{fig:12}(a).   

\begin{figure}
\includegraphics[width=3.5in]{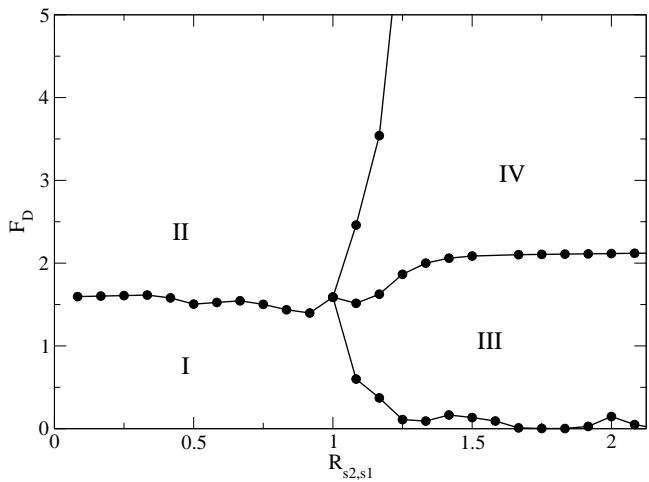}
\caption{ 
The three channel dynamic phase diagram for 
$F_{D}$ vs $R_{s2,s1}$ in the system 
from Fig.~\ref{fig:12} with $R_{s1,p}=1.0$.
The locations of regions I, II, III, and IV are marked.
Peaks appear in the value of $F_D$ at which region I ends for
the commensurate ratios of $R_{s2,s1} = 1.0$ and 
$R_{s2,s1}=2.0$
}
\label{fig:13}
\end{figure}

In Fig.~\ref{fig:13} we map out the dynamic phase diagram 
for a three channel system with $R_{s1,p} = 1.0$ and varied $R_{s2,s1}$.
The value of $F_D$ at which a transition out of region I occurs
shows commensurate peaks at $R_{s2,s1} = 1.0$ and 
$R_{s2,s1}=2.0$, while 
the region III-region IV transition
falls at a  roughly constant value of $F_{D} = 2.1$. 
The region II-region IV transition line shifts to slightly
higher $F_D$ with increasing $R_{s2,s1}$. This trend continues for
$F_D$ values higher than those shown in Fig.~\ref{fig:13}, until
at $R_{s2,s1}=2.0$ the II-IV transition drops to a value of
$F_D=7.5$ (not shown in the figure).
These results indicate that commensurability effects also occur
in the moving phases at high $F_D$.

\begin{figure}
\includegraphics[width=3.5in]{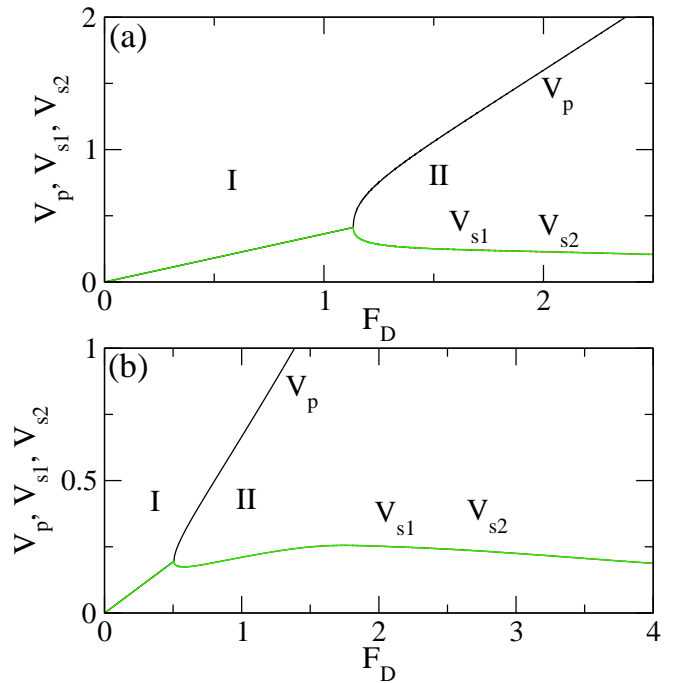}
\caption{ 
$V_p$, $V_{s1}$, and $V_{s2}$ versus
$F_D$ for a three channel system with $R_{s1,p} = 0.75$ and
varied $R_{s2,p}$.
(a) At $R_{s2,p} = 1.0$ there is a single transition from 
region I to region II.
(b) At 
$R_{s2,p} = 0.833$, 
the single region I-region II transition is
accompanied by an additional
secondary maximum in $V_{s1}$ 
and $V_{s2}$ centered at $F_D=1.75$.
}
\label{fig:14A}
\end{figure}

\begin{figure}
\includegraphics[width=3.5in]{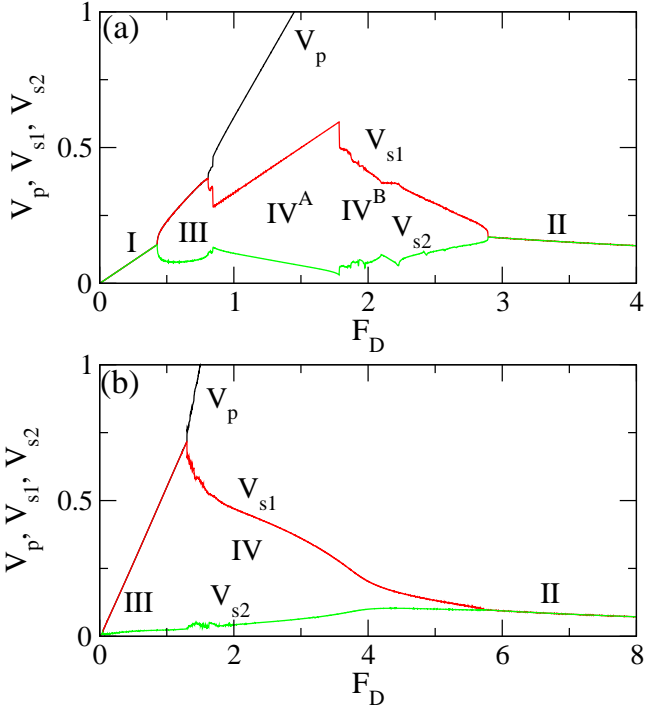}
\caption{ 
$V_p$, $V_{s1}$, and $V_{s2}$ versus
$F_D$ for a three channel system with $R_{s1,p} = 0.75$ and
varied $R_{s2,p}$.
(a) At $R_{s2,p} = 1.25$, region I is followed by a transition into
region III.  
In region IV$^{A}$, all the
channels are unlocked, $V_{s1}$ increases with increasing $F_D$,
and $V_{s2}$ decreases with increasing $F_D$. 
In region IV$^{B}$,  $V_{s1}$ decreases 
with increasing $F_D$ and $V_{s2}$ increases 
with increasing $F_D$. There is a transition to region II
at high $F_D$. 
(b) At $R_{s2,p} = 1.583$, there is a small window 
of region I at low $F_{D}$ which
is not highlighted on the figure. 
There is a transition directly from region III to 
region IV$^{B}$, with region IV$^A$ absent.
}
\label{fig:14B}
\end{figure}

\subsection{Dynamics for Increased Incommensuration}

In Figs.~\ref{fig:14A} and \ref{fig:14B} 
we plot $V_p$, $V_{s1}$, and
$V_{s2}$ versus $F_D$ for 
a three channel system with 
$R_{s1,p} = 0.75$ and varied $R_{s2,p}$. 
At $R_{s2,p} = 1.0$, shown in Fig.~\ref{fig:14A}(a), there
is a single decoupling transition from region I to region II 
at $F_{D} = 1.14$. For $R_{s2,p} < 1.0$, 
only regions I and II occur; however, 
$V_{s1}$ and $V_{s2}$
may contain additional features such as 
those shown in Fig.~\ref{fig:14A}(b) for 
$R_{s2,p} = 0.833$.
Here the decoupling into region II 
occurs near $F_{D} = 0.5$ which is significantly
lower than the location of the I-II transition in
the $R_{s2,p} = 1.0$ case. 
There is also a
secondary maximum 
in $V_{s1}$ and 
$V_{s2}$ near $F_{D} = 1.7$ which is similar to the secondary
maximum that appears in $V_{s1}$ at incommensurate fillings in 
the two channel system.  
At $R_{s2,p} = 1.25$ in Fig.~\ref{fig:14B}(a), 
the sample first transitions 
at $F_D=0.8$ from region I to region III, where the
particles in $s_1$ and $p$ are locked but the particles in $s_2$ are
partially decoupled.
At $F_D=0.86$ the sample enters region IV 
where all the channels are unlocked. 
The III-IV transition is also marked by a change in sign of the slope
of the velocity-force curves for both of the secondary channels. 
In Fig.~\ref{fig:14B}(a) we divide region IV into two subregions. 
Just above the III-IV transition we have region IV$^{A}$ in which
$V_{s1}$ increases
with increasing $F_D$ while $V_{s2}$ decreases 
with increasing $F_D$.
In region IV$^B$ this behavior is reversed and $V_{s1}$ 
decreases while $V_{s2}$ increases with increasing $F_D$.
When the particles in $s_1$ and $s_2$ recouple, a transition from
region IV$^B$ to region II occurs and both
$V_{s1}$ and $V_{s2}$ decrease with
increasing $F_D$.
There are two distinct subregions of region IV 
only for $1.0 < R_{s2,p} < 1.5$.  For $R_{s2,p} > 1.5$, only 
region $IV^{B}$ appears, as shown in Fig.~\ref{fig:14B}(b) 
for $R_{s2,p} = 1.583$ where the III-IV$^B$ transition falls at $F_D=1.3$.
There is a small window of region I that occurs at very
low $F_D$ which is not highlighted in the figure. 

\begin{figure}
\includegraphics[width=3.5in]{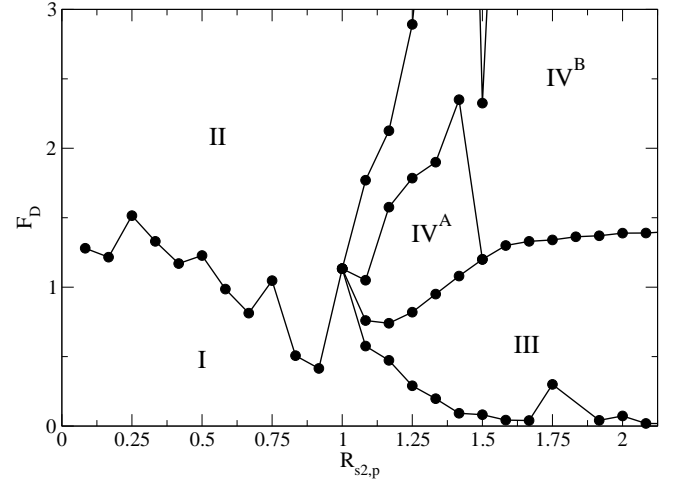}
\caption{ 
The dynamic phase diagram of $F_{D}$ vs $R_{s2,p}$ for the system in
Fig.~\ref{fig:14A} with $R_{s1,p}=0.75$.
}
\label{fig:15}
\end{figure}

Figure~\ref{fig:15} 
shows the $F_D$ versus $R_{s2,p}$ phase diagram 
for the system in Fig.~\ref{fig:14A} with $R_{s1,p}=0.75$ and $d/a=0.67$.
Here, the value of $F_D$ at which region I ends passes through peaks at
the commensurate values of 
$R_{s2,p} = 0.25$, 0.5, 0.75, 1.0, and 1.75, with 
a weaker peak at $R_{s2,p}=2.0$. 
The pronounced peak at $R_{s2,p} = 0.75$ 
also corresponds to the commensurability condition $R_{s2,s1} = 1.0$.
Region IV$^A$ first appears for $R_{s2,p}=1.0$ and 
vanishes at $R_{s2,p} = 1.5$, which is the $R_{s2,s1} = 2.0$ filling.
The value of $F_{D}$ at which the transition from region II to 
region IV$^B$ occurs increases with increasing $R_{s2,p}$, except at
$R_{s2,p}=1.5$ where the II-IV$^B$ transition suddenly drops to a lower
value of $F_D$.

\begin{figure}
\includegraphics[width=3.5in]{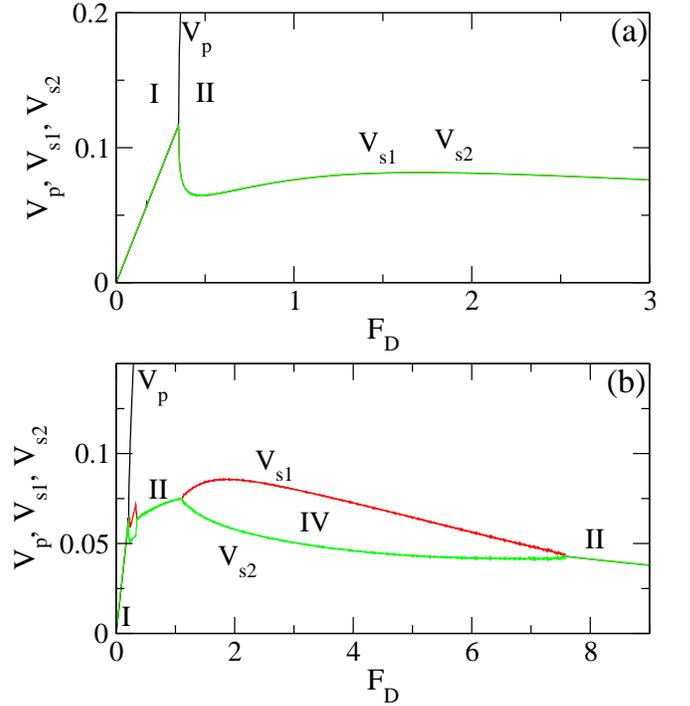}
\caption{ 
$V_p$, $V_{s1}$, and $V_{s2}$ vs
$F_D$
for the three channel system with $R_{s1,p} = 1.25$ and $d/a=0.67$.
(a) At $R_{s2,s1} =  0.6$ only regions I and II are present.
(b) At $R_{s2,s1} = 0.86$ the system enters region II more than
once.
}
\label{fig:16}
\end{figure}

In Fig.~\ref{fig:16}(a) we plot
$V_p$, $V_{s1}$, and $V_{s2}$
versus $F_D$ for a three layer system with $R_{s1,p} = 1.25$ at 
$R_{s2,s1} = 0.6$.
Only regions I and II are present, and there is an additional
second broad maximum in 
$V_{s1}$ and $V_{s2}$ 
centered near $F_{D} = 1.5$. 
In general, for $R_{s1,p}=1.25$ 
and $R_{s2,s1} < 0.8$ or $R_{s2,s1} > 1.0$, only regions I and II
appear.   
For $0.86 \leq  R_{s2,s1} < 1.0$, region II is
broken into two sections by an
intermediate transition to region IV,
as shown in Fig.~\ref{fig:16}(b) for $R_{s2,s1}=0.86$. 
The system passes from region I to region II, then enters
region IV and finally returns to region II at high $F_D$.

\begin{figure}
\includegraphics[width=3.5in]{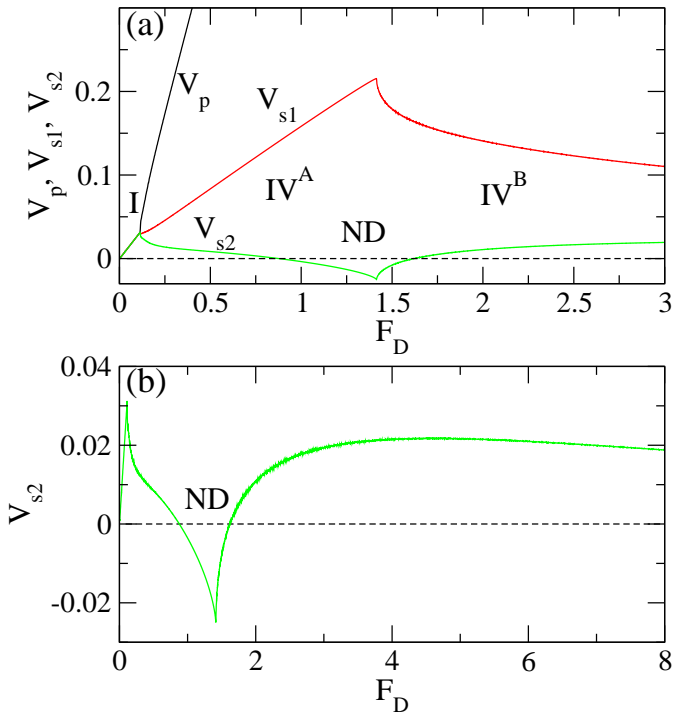}
\caption{ 
$V_p$, $V_{s1}$, and $V_{s2}$ versus
$F_D$ for the three channel system
with $R_{s1,p} = 1.25$.
(a) At $R_{s2,s1} = 1.067$, there is 
a transition from the locked region I to 
region IV$^A$.  This is followed by a transition to region IV$^B$. 
$V_{s2}$ drops below zero in the region marked ND where
negative drag occurs.
(b) $V_{s2}$ vs $F_D$ from (a) showing the region of negative
velocities as well as the existence of a local maximum at higher $F_D$.
}
\label{fig:17}
\end{figure}

The dynamic phase diagrams presented here give a concise description of the
velocity-force curves as the system parameters are varied.  Such 
dynamic phase diagrams have been widely used in studies of driven particle
systems such as vortices in type-II superconductors \cite{Periodic}; however,
they have no connection with equilibrium phase diagrams obtained from systems
in the thermodynamic limit.  Having more than three phase transition lines
meet in an equilibrium phase diagram would be highly unusual; however, in 
the nonequilibrium dynamic phase diagram, having more than three lines meet
has no special implications since the lines do not represent true
phase transition lines.
Whether nonequilibrium systems can undergo true phase  
transitions that resemble equilibrium phase transitions 
is currently a topic of active study and 
is beyond the scope of this manuscript to address. 
The appearance of multiple phases typically
occurs when the $s_1$ and $s_2$ channels can become unlocked with each other. 
For $R_{s1,s2} < 1.0$,
incommensurations in the form of holes are present in one channel;
however, the mobility of the holes is less than that
of the interstitials which arise when $R_{s1,s2} > 1.0$.       

\subsection{Negative Drag}  
For $ 1.0 < R_{s2,s1} \leq 1.6$ 
and fixed $R_{s1,p} = 1.25$, we show that a negative drag effect
can occur for the particles in $s_2$. 
During negative drag, the particles in $s_2$ move in the direction opposite
to the direction in which the particles in $p$ are being driven.
Negative drag has been observed in 
coupled 1D wires where Wigner crystallization is 
expected to occur \cite{Stopa}. 
In Fig.~\ref{fig:17}(a) we plot $V_p$, $V_{s1}$, and
$V_{s2}$ for a three channel system with 
$R_{s2,s1} = 1.067$. 
Here the sample is in the locked region I for $F_{D} < 0.1$. 
For $ 0.1 < F_{D} < 1.4$, region IV$^{A}$ appears with all three channels
decoupled, 
$V_{s1}$ increasing with increasing $F_D$, and
$V_{s2}$ decreasing with increasing $F_D$.
At $F_{D} = 1.4$ there is a cusp in both 
$V_{s1}$ and $V_{s2}$ at the onset
of region IV$^B$. 
The cusp also marks the point at which $V_{s2}$ reaches
its maximum negative value.  In Fig.~\ref{fig:17}(a) this is labeled ND
for the negative drag region, which extends from 
$1.0 < F_D < 1.6$.
In Fig.~\ref{fig:17}(b) we plot $V_{s2}$ alone 
versus $F_{D}$ for the system in 
Fig.~\ref{fig:17}(a) showing the negative drag effect more clearly and 
also showing the presence of a local maximum in $V_{s2}$
at $F_{D} = 4.5$.  Above this drive, 
$V_{s2}$ decreases with increasing $F_D$ 
but remains positive. 

\begin{figure}
\includegraphics[width=3.5in]{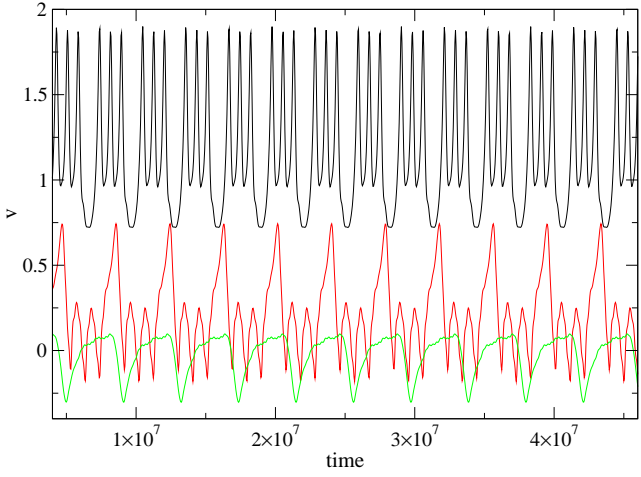}
\caption{ 
The time dependent velocity $v$ of a single particle in each of the 
channels for the system 
in Fig.~\ref{fig:16} at $F_{D} = 1.36$. 
Upper curve: $p$; middle curve: $s_1$; lower curve: $s_2$.
The velocity of the particle in $p$ exhibits
two frequencies and is always positive.  The velocity 
of the particle in $s_1$ also shows two frequencies and passes below
zero for a portion of each cycle, but the time averaged velocity
remains positive.
The particle in $s_2$ spends a larger fraction of each cycle moving in the
negative direction, producing a negative time averaged velocity.  
}
\label{fig:18}
\end{figure}

In Fig.~\ref{fig:18} we plot 
the time dependent velocity $v(t)$ 
of a single particle in each of of the three channels for the 
system in Fig.~\ref{fig:17} at $F_{D} = 1.36$ where the particles in $s_2$
undergo negative drag.
The velocity of the particle in $p$ is always positive 
and is composed of two frequencies. 
The velocity of the particle in $s_1$ again shows two frequencies
and drops below zero for a portion of each cycle; however, 
the overall time average of the velocity remains
positive. 
The particle in $s_2$ also experiences a 
combination of positive and negative velocities; however, the negative
velocity portion of each cycle is greater than the positive 
velocity portion, and
the particle takes a step backwards at the negative cusp in each cycle.
It was previously demonstrated
that a system driven by two external ac drives
can exhibit a ratchet effect
in the absence of an asymmetric substrate 
\cite{Hastings2,Libal,Ren}.  
In our three channel system, 
when $N_p$, $N_{s1}$, and $N_{s2}$ are all different,
the dynamical potential produced by the particles
in $p$ and $s_1$ acts effectively like two ac driving signals for
the particles in $s_2$.
In some cases, 
the interfering
frequencies of these ac drives 
can create a local potential maximum in $s_2$ that is moving in a direction
opposite to $F_D$.
As $F_{D}$ is further increased, the different
ac frequencies shift,  
increasing or decreasing the ratchet effect 
until for high enough $F_{D}$ the coupling between $s_2$ and
the particles in the other channels becomes so weak that a ratchet
effect can no longer occur.

\begin{figure}
\includegraphics[width=3.5in]{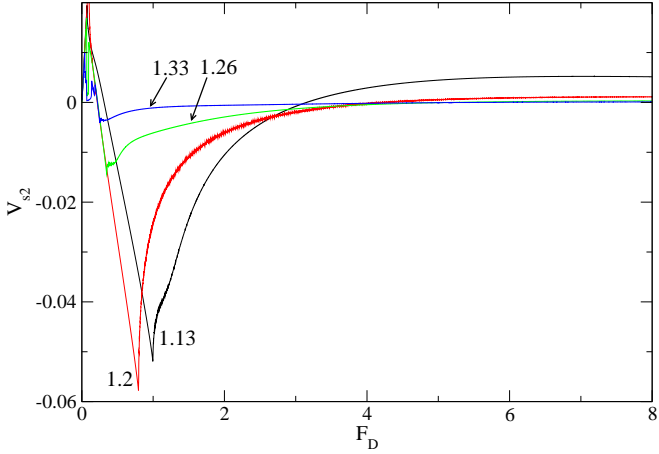}
\caption{ 
$V_{s2}$ vs $F_{D}$ 
for the system in Fig.~\ref{fig:16} at 
$R_{s2,s1}=1.13$, 1.2, 1.26, and 1.33, as labeled.
The largest negative maximum occurs for $R_{s2,s1}=1.2$.
}
\label{fig:19}
\end{figure}

\begin{figure}
\includegraphics[width=3.5in]{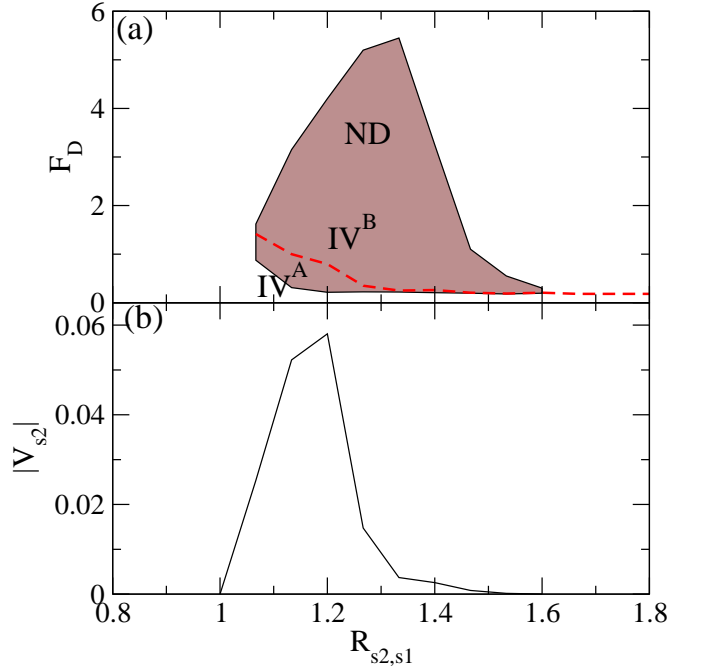}
\caption{ 
(a) $F_{D}$ vs $R_{s2,s1}$ for the system in Fig.~\ref{fig:16}.  The shaded
region marked ND indicates where 
the negative drag for the particles in $s_2$ occurs. 
Dashed line: The transition between regions
IV$^{A}$ and IV$^{B}$.  The largest negative maximum of 
$V_{s2}$ falls on this line. 
(b) $|V_{s2}|$, the magnitude of the largest negative
maximum in $V_{s2}$ in the negative drag region, vs $R_{s2,s1}$
for the same system. 
}
\label{fig:20}
\end{figure}

In Fig.~\ref{fig:19} we plot only the normalized velocities 
$V_{s2}$ versus 
$F_{D}$ for three channel samples with $R_{s1,p}=1.25$ and
$R_{s2,s1} = 1.13$, 1.2, 1.26, and $1.33$. 
This shows how the magnitude and extent of the negative
drag region changes with filling.
The negative velocity is maximum for $R_{s2,s1} = 1.2$ 
and gradually decreases with increasing $R_{s2,s1}$.   
In Fig.~\ref{fig:20}(a), the plot of $F_D$ versus $R_{s2,s1}$ is marked
with the region ND where negative drag occurs.
The dashed line indicates the location of the
the transitions between region IV$^{A}$ and  
region IV$^{B}$.  This transition also coincides with 
the maximum negative value of $V_{s2}$ 
for fixed $R_{s2,s1}$. 
In Fig.~\ref{fig:20}(b) we show 
$|V_{s2}|$ taken at the IV$^A$-IV$^B$ transition 
as a function of $R_{s1,s2}$, showing that the overall maximum
negative value of $V_{s2}$
occurs at $R_{s1,s2} = 1.2$. 

\begin{figure}
\includegraphics[width=3.5in]{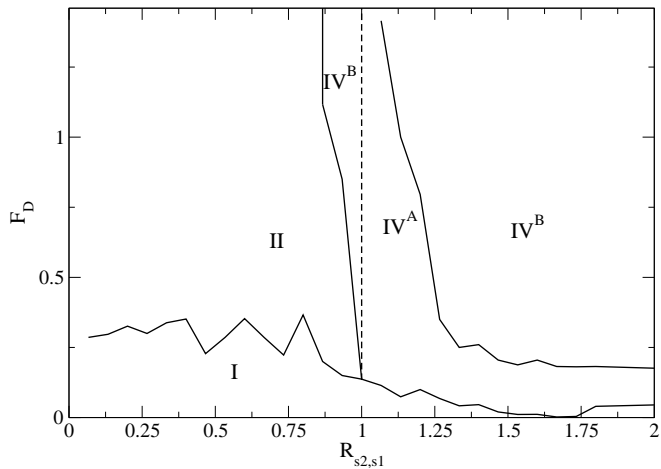}
\caption{ 
The dynamic phase diagram of $F_{D}$ vs $R_{s2,s1}$ for 
the three channel system with $R_{s1,p} = 1.25$ and $d/a=0.67$.
The prominent commensurate peak in the region I-region II transition at 
$R_{s2,s1} = 0.8$ also corresponds to the commensurability
condition of $R_{s2,p} = 1.0$. 
The dashed line indicates that at $R_{s2,s1} = 1.0$, the
system crosses from region IV$^B$ to region IV$^A$. 
At higher $F_D$ (not shown), the line marking the end of region IV$^A$
approaches $R_{s2,s1}$ from above, and once it reaches $R_{s2,s1}$,
region IV$^A$ disappears.
Also at higher $F_D$ (not shown), the line marking the beginning of
region IV$^B$ approaches $R_{s2,s1}$ from below, producing a transition
from region IV$^B$ to region II with increasing $F_D$.
}
\label{fig:21}
\end{figure}

In Fig.~\ref{fig:21} we show the dynamic phase diagram of 
$F_{D}$ versus $R_{s2,s1}$ for the three channel system with $R_{s1,p}=1.25$ and
$d/a=0.67$.
There are peaks in the transition out of region I at
$R_{s2,s1} = 0.4$, 0.6 0.8, 1.2, and $1.8$. 
These peaks correspond to 
$R_{s2,p} = 0.5$, 0.75, 1.0, 1.5, and $2.25$, 
with the most prominent peak appearing at $R_{s2,p} = 1.0$.
For $R_{s2,s1} < 0.8$ the system exhibits only regions I and II,
while for $0.8 < R_{s2,s1} \leq 1.0$, the transition from region I to
region II is followed by a transition into region IV$^B$ at higher
$F_D$.  At even higher $F_D>7.7$, not shown in the figure, the line marking
the transition from region II to region IV$^B$ changes curvature and
approaches $R_{s2,s1}=1.0$ with increasing $F_D$.  As a result, 
for $0.8 < R_{s1,s2} \leq 1.0$
there is
a high-drive transition from region IV$^B$ back to region II (not shown) when
the particles in $s_1$ and $s_2$ recouple, 
similar to the region IV-region II transition illustrated at high $F_D$
in Fig.~\ref{fig:16}(b).
At $R_{s2,s1} = 1.0$, the dashed line indicates the transition
from region IV$^B$ to region IV$^A$.
For $R_{s2,s1}>1.5$, the upper
region IV$^A$-region IV$^B$ transition saturates to the line $F_D=0.18$.
Near $R_{s2,s1}=1.0$, the upper IV$^A$-IV$^B$ transition
line approaches $R_{s2,s1}$ from above with increasing $F_D$, 
and when the transition reaches
$R_{s2,s1}$ below $F_D=2$ (not shown in the figure), region IV$^A$ disappears.
For $R_{s2,s1}>1.0$, we find no recoupling transition back into region II 
within the range $F_D \leq 15.0$.
Additionally, region III, where the particles in $p$ and $s_1$ are locked
but the particles in $s_2$ are unlocked, never occurs at all.
We have performed additional simulations 
for varied $R_{s1,p}>1.0$ other than the value $R_{s1,p}=1.25$ shown
in Fig.~\ref{fig:21} and find that the same sequence of regions
illustrated in the figure appears in each case.

\begin{figure}
\includegraphics[width=3.5in]{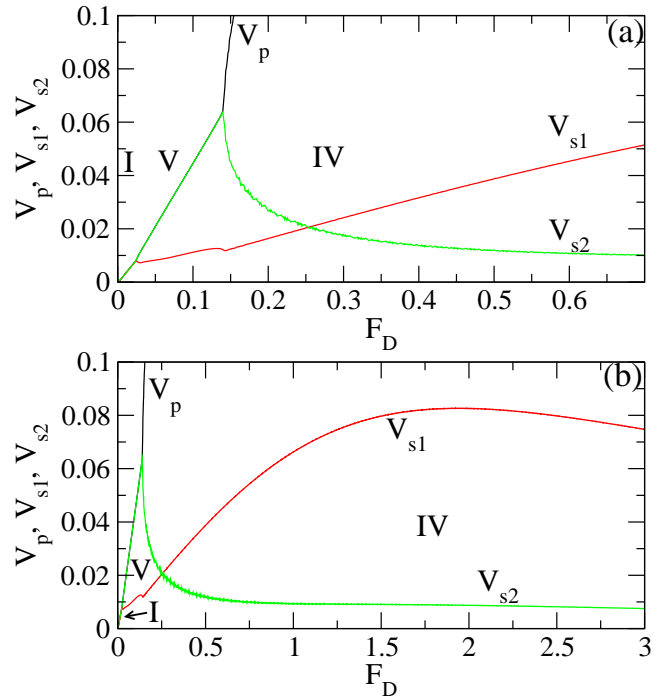}
\caption{
$V_p$, $V_{s1}$, and $V_{s2}$ vs $F_D$ for
a three channel system with $R_{s1,p}=R_{s2,s1}=1.133$, $R_{s2,p}=1.0$, and
$d/a=0.94$. 
Here we observe a transition from region I to 
region V, where the particles in $p$ and $s_2$ remain locked
to each other but the particles in $s_1$ are unlocked.
This is followed by region IV, when the particles in $s_2$ unlock from the
particles in $p$ and $V_{s1}$ increases with increasing
$F_D$.
(b) The same data plotted over a larger
range of $F_{D}$ shows that $V_{s1}$ 
reaches a plateau at $F_D=1.9$ and then decreases with
increasing $F_D$.       
}
\label{fig:22}
\end{figure}

\subsection{Unlocking of the Central Channel}
Another possible dynamic phase has the particles in $p$ and $s_2$ locked
with each other while the particles in $s_1$ are unlocked.  We term this
region V, and expect it to occur when the average interaction between the
particles in $p$ and $s_2$ is greater than the interaction between
the particles in $p$ and $s_1$ even though the distance between $s_1$ and
$p$ is shorter than the distance between $s_2$ and $p$.
In Fig.~\ref{fig:22}(a) we show an example of 
the occurrence of region V in a system with 
$R_{s1,p} = 1.133$, $R_{s2,s1} = 1.133$, 
$R_{s2,p} = 1.0$, and 
$d/a = 1.06$. 
In this case the $p$ and $s_2$ channels are commensurate.  
At low $F_D$, the system is in the locked phase I.  As $F_D$ increases,
the particles in $s_1$ decouple from the particles in $s_2$ and $p$, which
remain locked to each other.  This is indicated by the region in which
$V_{s1}$ splits away from $V_p$ and $V_{s2}$ and
increases at a diminished rate with increasing $F_D$.
At $F_{D} = 0.14$, the particles in $s_2$ also decouple from $p$ and
the system enters region IV, in which
$V_{s2}$ monotonically decreases with increasing $F_D$.
After the particles in $s_2$ decouple from the particles in $p$, the coupling
between the particles in $p$ and $s_1$ is increased, as indicated by the
increase in the slope of $V_{s1}$ at the onset of 
region IV.
$V_{s1}$ continues to increase 
with increasing $F_D$ throughout region IV
and even rises above $V_{s2}$ for $F_{D} > 0.2$. 
In Fig.~\ref{fig:22}(b) we plot the same data
over a larger range of $F_{D}$ 
to show that $V_{s2}$ 
reaches a maximum value near $F_{D}=1.9$ before turning over and beginning
to decrease with increasing $F_D$.

The results in Fig.~\ref{fig:22}
show that it is possible to achieve region V in certain situations, such
as when the particles in $p$ and $s_2$ are commensurate.
In general it is very difficult to obtain region V behavior in our system.
The coupling between the particles in $p$ and those in $s_2$ is relatively
weak since the distance between $p$ and $s_2$ is equal to
the screening length.  As a result, particles in $s_2$ experience a weak
interaction only with those 
particles in $p$ that lie directly above their positions,
and interact much more weakly still with the other particles in $p$.
(Note that we do not cut off the interaction at the screening length, but
continue to compute the weak interaction out to longer distances.)
This suggests that for different screening lengths $1/\kappa$ and
interchannel distances $d$
the coupling between particles in $p$ and particles in $s_2$ could be
enhanced, producing a more widespread occurrence of region V and leading
to additional commensuration effects.
The densities of the particles in the channels, and not merely the ratio of
their numbers, also plays an important role in determining which dynamical
regions will appear.
For higher particle density (smaller $a$), the couplings between the particles
in all the channels are reduced, as demonstrated for the
two channel case in Fig.~\ref{fig:3}(b).
Even if the effective coupling between the 
particles in $p$ and those in $s_2$ is strengthened by altering the
density of the particles in the channel, this coupling must still be stronger
than the coupling between the particles in $p$ and those in $s_1$ in order for
region V to appear. 

\begin{figure}
\includegraphics[width=3.5in]{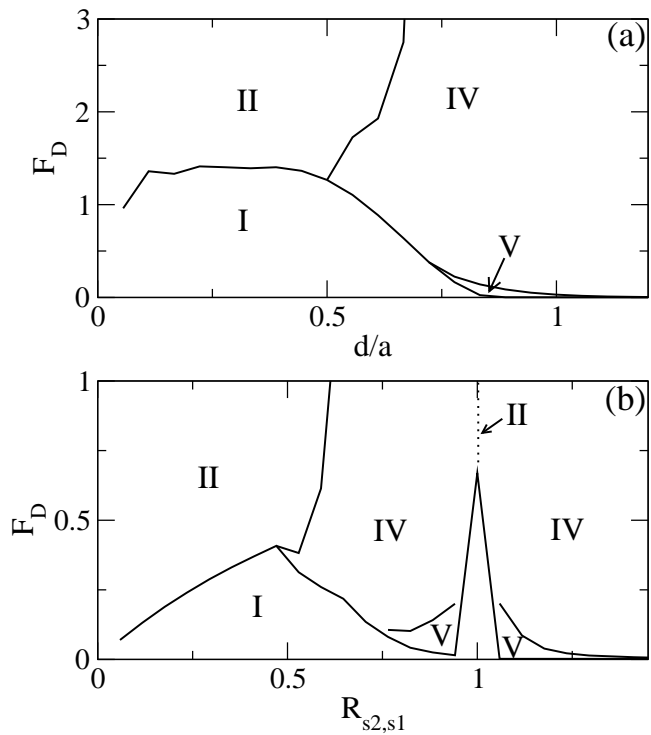}
\caption{ 
(a) The dynamic phase diagram of $F_{D}$ vs $d/a$ 
for a system with $R_{s1,p}=1.133$, $R_{s2,p} = 1.0$,
and $R_{s2,s1}=0.883$.
Here region V occurs for $d/a > 0.75$. 
(b) The dynamic phase diagram of 
$F_{D}$ vs $R_{s2,s1}$ for a system with $R_{s2,p}=1.0$ and
$d/a = 1.06$.
Commensurability peaks appear in the transition out of region I at
$R_{s2,s1} = 1.0$ and $R_{s2,s1}=0.5$.  
Region V appears on either side of the commensuration peak 
at $R_{s2,s1}=1.0$. 
The dashed line indicates that at $R_{s2,s1} = 1.0$, 
the system passes directly from region I to region II.
}
\label{fig:23}
\end{figure}

In order to understand where region V 
occurs as a function of the coupling between the channels, in
Fig.~\ref{fig:23}(a) we plot the dynamic 
phase diagram of $F_{D}$ versus $d/a$ for a system with 
fixed 
$R_{s1,p} = 1.133$, 
$R_{s2,p}=1.0$, and 
$R_{s1,s2} =  0.883$. 
For small $d/a$ the system passes directly from region I to region II.
At $d/a = 0.5$ a window of region IV opens between regions I and II.
Region V first appears at $d/a = 0.75$, and
gradually disappears for increasing $R_{s2,s1}$.
We next consider the case of 
$d/a=1.06$ and $R_{s2,p} = 1.0$ for varied $R_{s2,s1}$, 
as shown in Fig.~\ref{fig:23}(b). 
Here, there is a pronounced commensurability peak in the transition out of
region I at $R_{s2,s1} = 1.0$, where all the channels contain the
same number of particles. 
At $R_{s2,s1}=1.0$ the system passes directly
from region I to region II. 
In windows just below and just above $R_{s2,s1} = 1.0$ 
we find that region V appears and is accompanied by a transition to region IV
with increasing $F_D$.
The width of region V grows as $R_{s2,s1}=1.0$ is approached
from either side. 
For $R_{s2,s1} > 1.0$, region V gradually decreases in size with increasing
$R_{s2,s1}$, while for $R_{s2,s1} < 0.75$, region V vanishes completely. 
For $0.5 < R_{s2,s1} < 0.75$ the system transitions from region I into
region IV with increasing $F_D$ and eventually enters region II at high
$F_D$ (not shown).
For $R_{s2,s1}< 0.5$ there is a only a single transition
from region I to region II. 
A second commensurate peak in the transition out of region I appears
at $R_{s2,s1} = 0.5$.         

\begin{figure}
\includegraphics[width=3.5in]{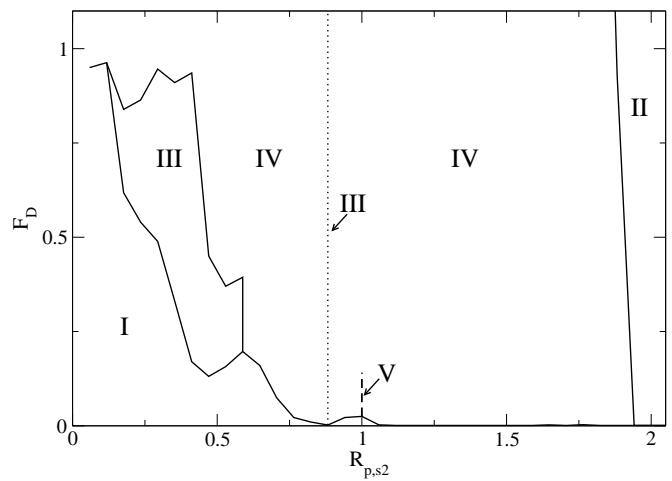}
\caption{ 
The dynamic phase diagram of $F_{D}$ vs $R_{p,s2}=N_p/N_{s2}$ 
for a system in which $N_p$
is varied.
Here 
$R_{s2,s1}=1.133$ and 
$d/a_{s1}=0.833$,
where $a_{s1}$ is the spacing of the particles in $s_1$.
All five regions appear as marked.  Region IV can be subdivided into
regions IV$^A$ and IV$^B$ as discussed previously, 
but for clarity this subdivision is omitted here.
Commensuration peaks at the transition out of region I
appear at $R_{p,s2} = 1.0$ and at $R_{p,s2} = 0.58825$; the latter  
corresponds to the commensurability condition of $R_{s1,p} = 2/3$. 
The dotted line at $R_{p,s2} = 0.882$ 
corresponds to $R_{s1,p} = 1.0$ where the system transitions directly
from region I to region III.  
A transition from region III to region IV occurs for this filling at 
$F_D=1.25$ (not shown).
At $R_{p,s2} = 1.0$ the transition from region I to region V is
marked by the thick dashed line.
For this filling,
region V ends at $F_{D} = 0.141$ and 
is followed by region IV.  
The transition from region IV to region II at high $R_{p,s2}$ continues to
rise to higher values of $F_D$ as $R_{p,s2}$ decreases over a range of
$F_D$ larger than shown in the figure.
}
\label{fig:24}
\end{figure}

\subsection{Five Dynamical Phases}

In Fig.~\ref{fig:24} we plot the dynamic 
phase diagram of $F_{D}$ versus $R_{p,s2}=N_p/N_{s2}$ for a system which exhibits
all five phases as well as several regions where a negative drag effect
occurs.
Here we vary $N_p$ and fix 
$R_{s2,s1}=1.133$ and 
$d/a_{s1}=0.833$, 
where $a_{s1}$ is the spacing
of the particles in $s_1$.
For this choice of parameters, we observe region V only 
at $R_{p,s2} = 1.0$, the value shown in Fig.~\ref{fig:22}.
At $R_{p,s2} = 0.882$, which also corresponds
to $R_{s1,p} = 1.0$, 
there is a single transition 
from region I to region III, indicated by the dashed line. 
Here the particles in $p$ and $s_1$
are locked because they are commensurate. 
The dynamics for $0.6 < R_{p,s2} < 1.75$ is dominated by region IV.
For $1.75 \leq R_{p,s2} < 1.95$, a transition from region IV to
region II occurs at higher $F_D$.  The location of this transition shifts to
higher values of $F_D$ as $R_{p,s2}$ drops below $R_{p,s2}=1.95$.  For
$R_{p,s2} \geq 1.95$, the system goes directly into region II for finite $F_D$.
Region II appears for high $R_{p,s2}$ since as $N_p$ increases, the effectiveness
of the coupling between the particles in $p$ and the particles in $s_1$ and
$s_2$ decreases.
As a result, even though the particles in $s_1$ and $s_2$ are incommensurate,
the coupling between the primary and secondary channels eventually becomes
so weak that the particles in $s_1$ and $s_2$ couple with each other
and decouple from the particles in $p$.
At $R_{p,s2} = 0.58825$ there is another peak in the transition out
of region I produced by the commensurability condition 
of $R_{s1,p} = 2/3$ at this filling. 
For $R_{p,s2} < 0.6$, region I grows in extent and there is a window of
region III which separates region I at low drives and region IV at higher
drives.
For $R_{p,s2} < 0.15$, there is a single transition from region I 
directly to region IV.  

\begin{figure}
\includegraphics[width=3.5in]{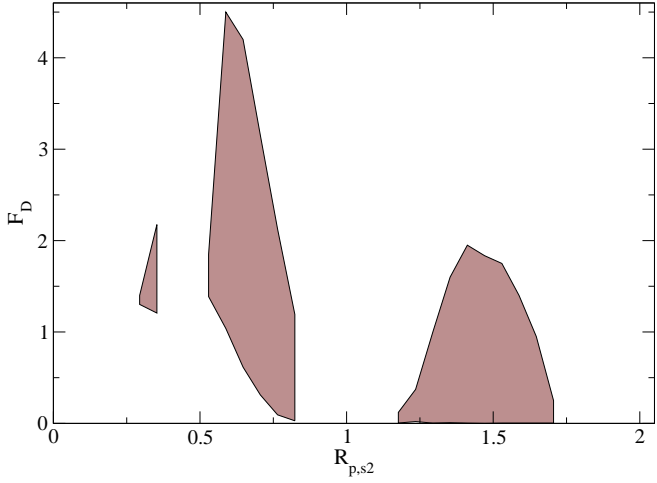}
\caption{ 
A plot of $F_D$ vs $R_{p,s2}$ for the system in Fig.~\ref{fig:25} indicating
the three regions in which negative drag of the particles in $s_2$ occurs.
}
\label{fig:25}
\end{figure}

\begin{figure}
\includegraphics[width=3.5in]{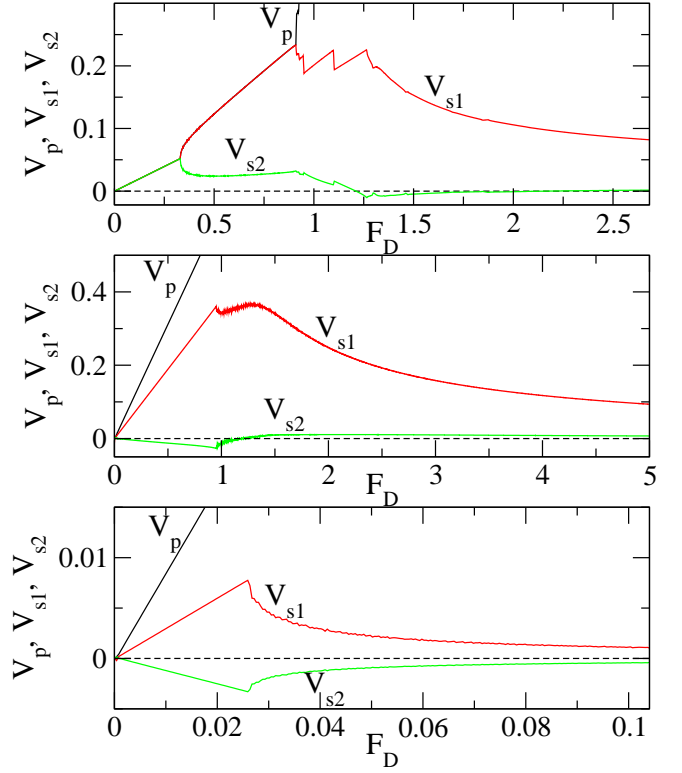}
\caption{ 
Representative velocity force curves 
$V_p$, $V_{s1}$, and $V_{s2}$ vs $F_D$ 
from each of the three regions where negative drive occurs
in Fig.~\ref{fig:25}. 
(a) $R_{p,s2} = 0.35$. (b) $R_{p,s2} = 0.823$.  
(c) $R_{p,s2} = 1.53$.
}
\label{fig:26}
\end{figure}

In Fig.~\ref{fig:25} we indicate the regions 
in the $F_{D}$ versus $R_{p,s2}$ plot where negative drag of the particles
in $s_2$ occurs
for the system in Fig.~\ref{fig:24}, and in 
Fig.~\ref{fig:26} we show representative velocity force curves 
for the three different negative drag regions. In 
Fig.~\ref{fig:25} the largest region
of negative drag occurs for $0.52 < R_{p,s2} < 0.82$. 
There is a small negative drag window near $R_{p,s2} = 0.3$. 
We illustrate a typical velocity force curve from this window
in Fig.~\ref{fig:26}(a) where we plot
$V_p$, $V_{s1}$, and $V_{s2}$
for $R_{p,s2} = 0.35$. Here the negative drag occurs in region IV. 
There are also a number of slip events which appear as
sharp changes in $V_{s1}$ near $F_{D} = 1.25$. 
For higher $F_{D}$ beyond what is shown in the 
figure, $V_{s2}$ continues to increase back above zero, passes
through a broad peak, and then slowly decreases back toward zero at high
$F_D$.
In Fig.~\ref{fig:26}(b) we plot the velocity force curves at
$R_{s2,p} = 0.823$ where the system exhibits only region IV flow.
Here the maximum negative value of $V_{s2}$ occurs at 
$F_{D}= 0.95$ in the form of a cusp which is accompanied by a cusplike peak
in $V_{s1}$. 
For $ 0 < F_{D} < 0.96$, $V_{s1}$ increases linearly with 
increasing $F_{D}$ but the particles in $s_1$ are not completely locked with the
particles in $p$.
This corresponds to region IV$^A$ as was discussed earlier; however, 
in the phase diagram of Fig.~\ref{fig:24} 
we omit the distinction between regions IV$^A$ and IV$^B$ for clarity. 
In Fig.~\ref{fig:26}(c) we plot the velocity force curves at 
$R_{p,s2} = 1.53$ in the third region of negative drag.  Here we find that
the magnitude of the maximum negative velocity in $V_{s2}$ is
reduced compared to the other two negative drag regions.

The general features of the phases outlined 
so far also occur for other parameters of density and
filling, indicating that they are robust features of the system. 
We have not observed negative drag of the particles in $s_1$ or $p$.

\section{Discussion and Summary}
We investigated a simple system consisting 
of two or three coupled 1D channels of particles interacting via a repulsive 
Yukawa potential where only one of the channels is driven. For two channel 
systems with an equal number of particles in each channel, we find 
a single transition from a completely locked state to 
a partially decoupled state where particles in the
secondary channel slip with respect to particles in the driven channel.
In the decoupled state, the velocity of the particles in the secondary channel
gradually decreases with increasing drive while the velocity of the driven
particles increases linearly with increasing drive.
When the number of particles in the secondary channel is 
slightly away from commensuration with the number of particles in the
primary channels,
a two stage decoupling transition occurs where 
the first decoupling is associated with individual slips of 
the incommensurations or vacancies in the secondary channel. 
The velocity of the particles in the secondary channel continues to
increase with increasing drive
until the second decoupling transition is reached, whereupon 
all the particles in the secondary
channel begin to slip.
The driving force at which the transition from the 
completely locked to the decoupled flow occurs has peaks
at integer commensurate ratios of the number of particles in the two channels
as well as at certain fractional ratios such as $1/2$ or $3/2$; 
however, there are no peaks for low filling ratios since the particles in
the driven channel are effectively moving not over a fixed substrate but over
a distortable substrate.
We also observe a ratchet effect in the two channel system
where the particles in the secondary channel can be rectified 
by an asymmetric ac drive applied to the primary channel. 
This ratchet effect is similar to the ratchet effect
found for coupled binary particle species where only one species is driven.  

For three channels we find that a remarkably rich variety       
of dynamical phases such as coupling and decoupling transitions
are possible and produce a variety of commensuration 
effects as well as pronounced signatures in the velocity force curves. 
The commensuration effects occur whenever the ratio of the 
number of particles in at least two of the channels is an integer or rational 
fraction.
We also observe a negative drag effect for the secondary channel which is
furthest from the driven channel.  Here, the particles in the secondary
channel move in the direction opposite to the driving direction of the
primary channel.
When the negative drag occurs, all three channels have incommensurate fillings.
The resulting multiple periodic forces experienced by the particle in the
furthest secondary channel create a
bi-harmonic ratchet effect of a type that has been observed in 
systems driven with multiple ac drives. 

Our results could be tested for colloidal particles 
confined to two or three channels where one of the
channels is driven by optical means or via microfluidics. 
Since the motion of physical colloids is never perfectly one-dimensional,
some smearing of the effects we observe might occur, but the general
features we describe should be observable.
A similar experiment could be performed in a dusty plasma system 
with the dust particles confined in grooves and driven in one dimension
with a laser focused in a single plane.
Some of the effects we observe could be relevant for certain 
superconducting vortex systems in which two different types of
vortices are coupled and one of the two vortex types is driven with an 
external current.
Additionally, these effects could also be realized using
coupled wires in which one-dimensional Wigner crystal
states occur.  The velocity-force responses that we predict 
could be a potentially powerful method for determining whether Wigner 
crystals are actually present in the wires. 
It would also be interesting to study ratchet effects 
with asymmetric ac drives for three or more channels.  Here, 
it may be possible to induce dc currents flowing in different 
directions for different channels.  
Although the system we consider appears very simple, 
we have shown that it exhibits a rich variety of behaviors even 
without substrates or other complications.  If a periodic substrate
were introduced in one or more of the channels, we expect that
an even greater variety of commensuration effects and 
coupling between excitations in the channels could occur. 
It would also be interesting to consider cases where 
the channels are not strictly one-dimensional but have
a finite width to allow for transitions 
to buckled or zig-zag states.  Even for the commensurate
fillings, such buckling transitions could
produce interesting new features in the drag behavior.  

This work was carried out under the auspices of the 
NNSA of the 
U.S. DoE
at 
LANL
under Contract No.
DE-AC52-06NA25396.


\begin{thebibliography}{99}

\bibitem{Koppi}
M.~K{\"o}ppl, P.~Henseler, A.~Erbe, P.~Nielaba, and P.~Leiderer,
Phys.~Rev.~Lett.~{\bf 97}, 208302 (2006);
P. Henseler, A.~Erbe, M.~K{\"o}ppl, P.~Leiderer, and P.~Nielaba,
Phys.~Rev.~E {\bf 81}, 041402 (2010). 

\bibitem{Mc}
D.~McGloin, A.E.~Carruthers, K.~Dholakia, and E.M.~Wright,
Phys.~Rev.~E {\bf 69}, 021403 (2004).

\bibitem{Doyl}
R.~Haghgooie and P.S.~Doyle,
Phys.~Rev.~E {\bf 72}, 011405 (2005). 

\bibitem{Ferreira}
W.P.~Ferreira, J.C.N.~Carvalho, P.W.S.~Oliveira, G.A.~Farias, and 
F.M.~Peeters, Phys.~Rev.~B {\bf  77}, 014112 (2008). 

\bibitem{Yang}
W.~Yang, K.~Nelissen, M.~Kong, Z.~Zeng, and F.M.~Peeters,
Phys.~Rev.~E {\bf 79}, 041406 (2009).

\bibitem{Misko}
D.V.~Tkachenko, V.R.~Misko, and F.M.~Peeters, Phys.~Rev.~E 
{\bf 80}, 051401 (2009).

\bibitem{Wei}
Q.-H.~Wei, C.~Bechinger, and P.~Leiderer,
Science {\bf 287}, 625 (2000).

\bibitem{Lutz}
C.~Lutz, M.~Kollmann, and C.~Bechinger,
Phys.~Rev.~Lett. {\bf 93}, 026001 (2004).

\bibitem{Bleil}
S.~Bleil, P.~Reimann, and C.~Bechinger,
Phys.~Rev.~E {\bf 75}, 031117 (2007).

\bibitem{Roichman}
Y.~Roichman, D.G.~Grier, and G.~Zaslavsky,
Phys.~Rev.~E {\bf 75}, 020401(R) (2007).

\bibitem{Sc}
H.J.~Schulz, Phys.~Rev.~Lett.~{\bf 71}, 1864 (1993).

\bibitem{Glaz}
L.I.~Glazman, I.M.~Ruzin, and B.I.~Shklovskii, 
Phys.~Rev.~B {\bf 45}, 8454 (1992).

\bibitem{R}
J.~Baker and A.G.~Rojo, J.~Phys.: Condens. Mat. {\bf 13}, 5313 (2001).

\bibitem{Mueller}
E.J.~Mueller, Phys.~Rev.~B {\bf 72},  075322 (2005). 

\bibitem{Bockrath}
V.V.~Deshpande and M.~Bockrath, Nature Phys. {\bf 4}, 314 (2008).

\bibitem{Hew}
W.K.~Hew, K.J.~Thomas, M.~Pepper, I.~Farrer, D.~Anderson, G.A.C.~Jones,
and D.A.~Ritchie, Phys.~Rev.~Lett. {\bf 102}, 056804 (2009).

\bibitem{Peeters1}
G.~Piacente, I.V.~Schweigert, J.J.~Betouras, and F.M.~Peeters,
Phys.~Rev.~B {\bf 69}, 045324 (2004);
G.~Piacente, G.Q.~Hai, and F.M.~Peeters, Phys.~Rev.~B {\bf 81}, 024108 (2010). 

\bibitem{I}
H.~Ikegami, H.~Akimoto, and K.~Kono, Phys.~Rev.~Lett. {\bf 102}, 046807 (2009).

\bibitem{Meyer}
J.S.~Meyer and K.A.~Matveev, J Phys.: Condens. Mat. {\bf 21}, 023203 (2009). 

\bibitem{FPeeters}
G.~Piacente and F.M.~Peeters, 
Phys.~Rev.~B {\bf 72}, 205208 (2005).

\bibitem{Goree}
B.~Liu, K.~Avinash, and J.~Goree, Phys.~Rev.~Lett. {\bf 91}, 
255003 (2003);
B.~Liu and J.~Goree, 
Phys.~Rev.~E {\bf 71}, 046410 (2005). 

\bibitem{Coupier}
G.~Coupier, M.~Saint Jean, and C.~Guthmann, 
Phys.~Rev.~E {\bf 73}, 031112 (2006);
C.~Coste, J.-B.~Delfau, C.~Even, and M.~Saint Jean, 
Phys.~Rev.~E {\bf 81},  051201 (2010).

\bibitem{Kes}
R.~Besseling, R.~Niggebrugge, and P.H.~Kes, Phys.~Rev.~Lett. {\bf 82}, 
3144 (1999); 
N.~Kokubo, R.~Besseling, V.M.~Vinokur, and P.H.~Kes,
Phys.~Rev.~Lett. {\bf 88}, 247004 (2002). 

\bibitem{K}
N.~Kokubo, T.G.~Sorop, R.~Besseling, and P.H.~Kes, 
Phys.~Rev. B {\bf 73}, 224514 (2006)

\bibitem{A}
P.~Barrozo, A.A.~Moreira, J.A.~Aguiar, and J.S.~Andrade,
Phys.~Rev.~B {\bf 80}, 104513 (2009).

\bibitem{Plourde}
K.~Yu, M.B.S.~Hesselberth, P.H.~Kes, and B.L.T.~Plourde, 
Phys.~Rev.~B {\bf 81}, 184503 (2010). 

\bibitem{Reichhardt}
C.J.~Olson Reichhardt and C.~Reichhardt, Phys.~Rev.~B {\bf 81}, 224516 (2010).

\bibitem{Ig}
I.~Giaever, Phys.~Rev.~Lett. {\bf 15}, 825 (1965).

\bibitem{Jr}
J.R.~Clem, Phys.~Rev.~B {\bf 9}, 898 (1974).

\bibitem{Ekin} 
J.W.~Ekin, B.~Serin, and J.R.~Clem, Phys.~Rev.~B {\bf 9}, 912 (1974) 

\bibitem{Pe}
T.~Pe, M.~Benkraouda, and J.R.~Clem,
Phys.~Rev.~B {\bf 56}, 8289 (1997).

\bibitem{Stopa}
M.~Yamamoto, M.~Stopa, Y.~Tokura, Y.~Hirayama, and S. Tarucha, 
Science {\bf 313}, 204 (2006); 
M.~Yamamoto, H.~Takagi, M.~Stopa, and S.~Tarucha, 
AIP Conf. Proc. {\bf 893}, 747 (2007).

\bibitem{Bak}
A review of commensurate-incommensurate transitions appears 
in P.~Bak, Rep.~Prog.~Phys. {\bf 45}, 587 (1982).

\bibitem{Cm}
S.N.~Coppersmith, D.S.~Fisher, B.I.~Halperin, P.A.~Lee, and W.F.~Brinkman,
Phys.~Rev.~B {\bf 25}, 349 (1982). 

\bibitem{Baert} 
M.~Baert, V.V.~Metlushko, R.~Jonckheere, V.V.~Moshchalkov, and Y.~Bruynseraede,
Phys.~Rev.~Lett.~{\bf 74}, 3269 (1995);
J.I.~Mart{\' \i}n, M.~V{\' e}lez, J.~Nogu{\' e}s, and I.K.~Schuller, 
Phys.~Rev.~Lett.~{\bf 79}, 1929 (1997);
S.~Avci, Z.L.~Xiao, J.~Hua, A.~Imre, R.~Divan, J.~Pearson, U.~Welp,
W.K.~Kwok, and G.W.~Crabtree, Appl.~Phys.~Lett.~{\bf 97}, 042511 (2010). 

\bibitem{Nori}
C.~Reichhardt, C.J.~Olson, and F.~Nori, Phys.~Rev.~B {\bf 57}, 7937 (1998);
G.R.~Berdiyorov, M.V.~Milosevi{\' c}, and F.M.~Peeters, 
Phys.~Rev.~B {\bf 74}, 174512 (2006).

\bibitem{Met}
M.~Baert, V.V.~Metlushko, R.~Jonckheere, V.V.~Moshchalkov, and Y.~Bruynseraede,
Europhys.~Lett. {\bf 29}, 157 (1995);
C.~Reichhardt and N.~Gr{\o}nbech-Jensen, Phys.~Rev.~B {\bf 63}, 054510 (2001).

\bibitem{Bechinger}
A.~Chowdhury, B.J.~Ackerson, and N.A.~Clark, 
Phys.~Rev.~Lett.~{\bf 55}, 833 (1985);
C.~Bechinger, M.~Brunner, and P.~Leiderer,
Phys.~Rev.~Lett. {\bf 86},  930 (2001).

\bibitem{Olson}
C.~Reichhardt and C.J.~Olson, Phys.~Rev.~Lett.~{\bf 88}, 248301 (2002);
M.~Brunner and C.~Bechinger, Phys.~Rev.~Lett. {\bf 88}, 248302 (2002);
K.~Mangold, P.~Leiderer, and C.~Bechinger, Phys.~Rev.~Lett.~{\bf 90}, 
158302 (2003). 

\bibitem{Periodic}
C.~Reichhardt, C.J.~Olson, and F.~Nori,
Phys.~Rev.~Lett.~{\bf 78}, 2648 (1997);
J.~Gutierrez, A.V.~Silhanek, J.~Van de Vondel, W.~Gillijns, 
and V.V.~Moshchalkov,
Phys.~Rev.~B {\bf 80}, 140514(R) (2009);
C.~Reichhardt and C.J.~Olson Reichhardt, Phys.~Rev.~B {\bf 81}, 024510 (2010). 

\bibitem{Hu}
O.M.~Braun, A.R.~Bishop, and J.~R{\" o}der, 
Phys.~Rev.~Lett.~{\bf 79}, 3692 (1997); 
J.~Teki{\' c}, O.M.~Braun, and B.~Hu, Phys.~Rev.~E {\bf 71}, 026104 (2005);
Y.~Yang, W.-S.~Duan, J.-M.~Chen, L.~Yang, J.~Teki{\' c}, Z.-G.~Shao,
and C.-L. Wang, Phys.~Rev.~E {\bf 82}, 051119 (2010).

\bibitem{Sav}
S.~Savel'ev and F.~Nori,
Nature Mater. {\bf 1}, 179 (2002).

\bibitem{Cole}
D.~Cole, S.~Bending, S.~Savel'ev, A.~Grigorenko, T.~Tamegai, and F.~Nori,
Nature Mater. {\bf 5}, 305 (2006).

\bibitem{Caplin}
M.~Tesei, G.K.~Perkins, A.D.~Caplin, L.F.~Cohen, and T.~Tamegai,
Supercond.~Sci.~Technol. {\bf 21}, 075019 (2008).

\bibitem{HL}
M.~Das, G.~Ananthakrishna, and S.~Ramaswamy, 
Phys.~Rev.~E {\bf 68}, 061402 (2003);
R.~Messina and H.~L{\" o}wen, Phys.~Rev.~E {\bf 73}, 011405 (2006). 

\bibitem{Olson2}
C.~Reichhardt and C.J.~Olson Reichhardt, Phys.~Rev.~E {\bf 79}, 061403 (2009).

\bibitem{Jiang}
Z.~Jiang, D.A.~Dikin, V.~Chandrasekhar, V.V.~Metlushko, and
V.V.~Moshchalkov, Appl.~Phys.~Lett. {\bf 84}, 5371 (2004).

\bibitem{Libal2}
A. Lib{\' a}l, C. Reichhardt, and C.J. Olson Reichhardt, 
Phys.~Rev.~Lett.~{\bf 97}, 228302 (2006);
A. Lib{\' a}l, C. Reichhardt, and C.J. Olson Reichhardt,
Phys.~Rev.~E {\bf 75}, 011403 (2007).

\bibitem{LowenN}
M. Rex and H. Lowen, Eur. Phys. J. E {\bf 26}, 143 (2008).

\bibitem{Reimann}
P.~Reimann, Phys.~Rep.~{\bf 361}, 57 (2002).

\bibitem{Hastings}
C.~Reichhardt, C.J.~Olson Reichhardt, and M.B.~Hastings,
Phys.~Lett.~A {\bf 342}, 162 (2005).

\bibitem{Hastings2}  
C.~Reichhardt, C.J.~Olson, and M.B.~Hastings,
Phys.~Rev.~Lett.~{\bf 89}, 024101 (2002);
R.~Guantes and S.~Miret-Art{\' e}s, Phys.~Rev.~E {\bf 67}, 046212 (2003);
C.~Reichhardt and C.J.~Olson Reichhardt, Phys.~Rev.~E {\bf 68},
046102 (2003);
V.~Lebedev and F.~Renzoni, Phys.~Rev.~A {\bf 80}, 023422 (2009);
D.~Speer, R.~Eichhorn, and P.~Reimann, 
Phys.~Rev.~Lett.~{\bf 102}, 124101 (2009). 

\bibitem{Libal}
A.~Lib{\' a}l, C.~Reichhardt, B.~Jank{\' o}, and C.J.~Olson Reichhardt, 
Phys.~Rev.~Lett.~{\bf 96}, 188301 (2006).

\bibitem{Ren}
L.~Machura and J.~Luczka, Phys.~Rev.~E {\bf 82}, 031133 (2010).  

\end{thebibliography}
\end{document}